# Strategic Plan for Astronomy in the Netherlands 2011-2020

Netherlands Committee for Astronomy
on behalf of
NOVA, SRON, ASTRON and NWO-EW

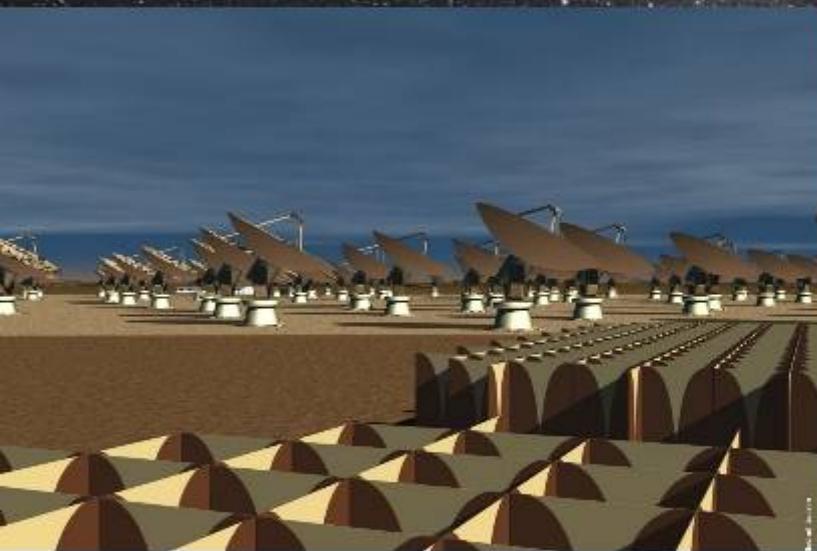
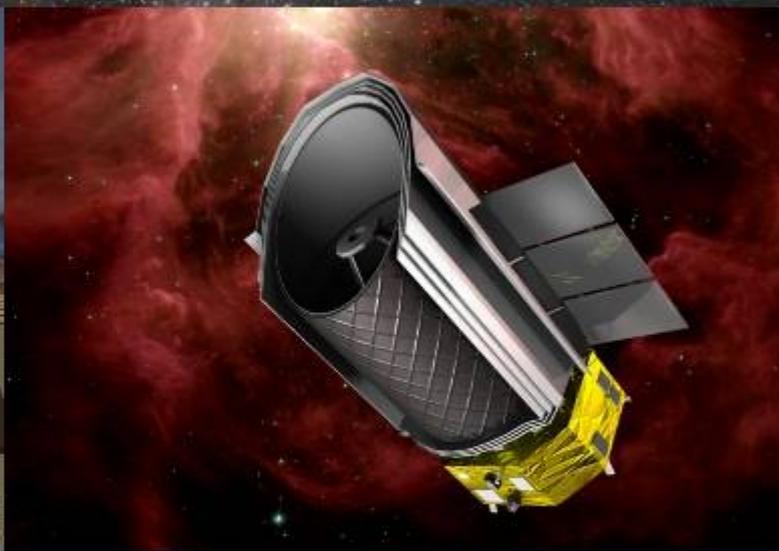
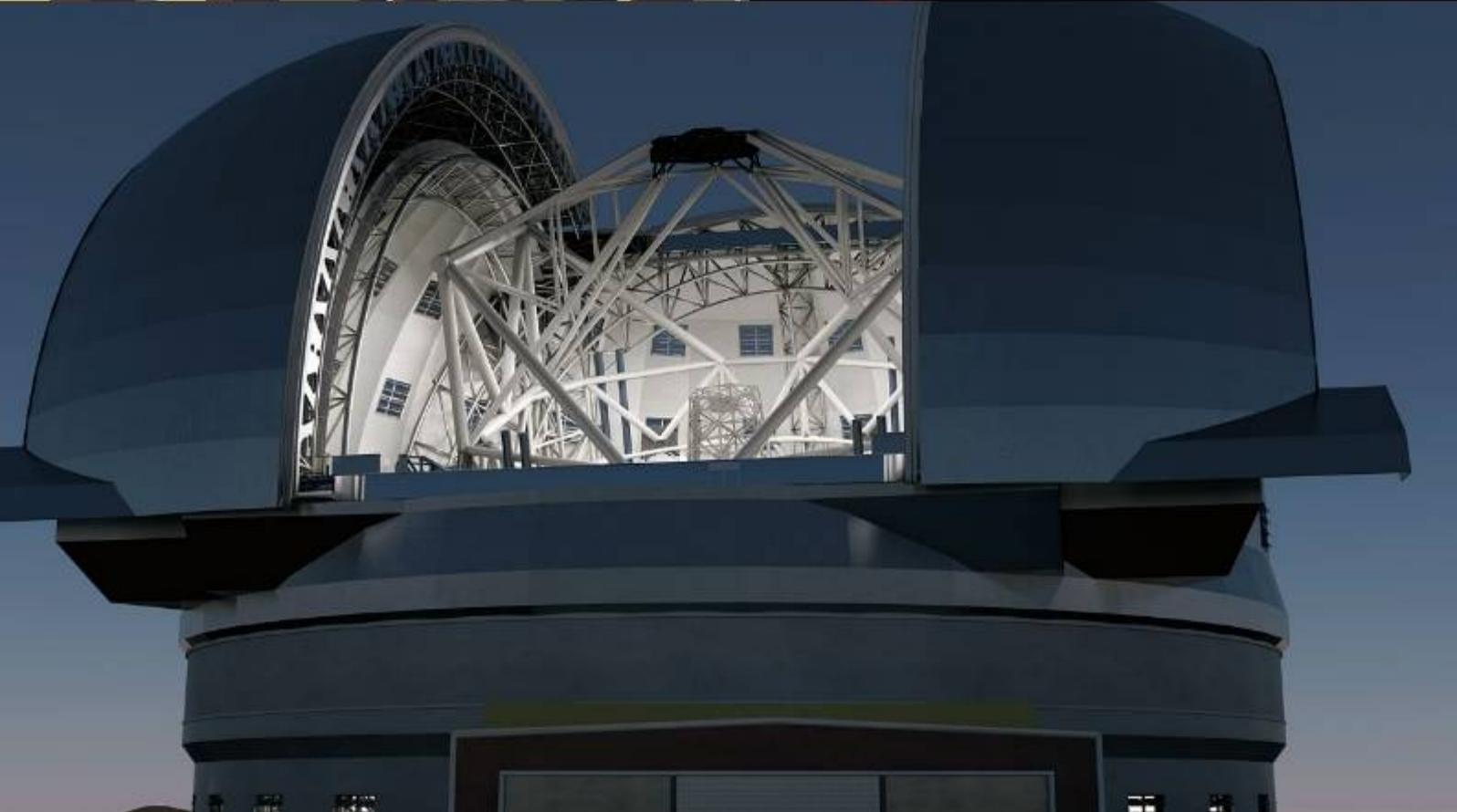

# Strategic Plan for Astronomy in the Netherlands 2011-2020

P.J. Groot & K. Kuijken, eds.

Netherlands Committee for Astronomy,
on behalf of NOVA, SRON, ASTRON
and NWO-EW

R.A.M.J. Wijers, chair
L. Kaper, secretary

| | |
|---|---|
| W. Boland | K. Kuijken |
| E. van Dishoeck | R.F. Peletier |
| M.A. Garrett | R. Stark |
| R. Gathier | F. Verbunt |
| P.J. Groot | M. de Vos |
| C.U. Keller | L.B.F.M. Waters |
| P.C. van der Kruit | |

19 June 2012

# Table of Contents

## Executive Summary

The remarkable progress in astronomy of the 20th century has revealed a strange and extremely exciting Universe. In a hot Big Bang the Universe started expanding about 13.7 billion years ago. Mysterious dark matter and dark energy make up 96% of its content and both will require new physics to explain their properties. Telescopes are now so powerful that the history of galaxy formation can be traced back to the earliest times, linking our own Milky Way Galaxy to the quantum fluctuations that led to all structure in the Universe. The processes by which gas clouds collapse and turn into stars and planets can be observed directly, including the building block of life. The Universe is a giant laboratory to study new physics under the most extreme conditions: near black holes and neutron stars, in the biggest explosions and just after the Big Bang.

Astronomy in the Netherlands is flourishing and Dutch astronomers play a leadership role in global astronomy. The focal points of Dutch astronomy are at the heart of the European ASTRONET roadmap. To maintain this vanguard position requires strong national support for training, attracting and keeping talent: top science is done by the brightest talents with the best tools in a mix of theory, numerical modelling and observations. Dutch astronomy must be at the forefront of designing, building, and exploiting the best instruments and facilities, world-wide, in close and natural partnership with neighbouring disciplines and industry, as part of the High-Tech infrastructure in the Netherlands. The inspiring big questions and their answers are to be shared with the general public and prospective new talents.

The conditions astronomers study in space are extreme, and their ongoing interpretations continually push the understanding of physics, chemistry and geology. Astronomy has motivated and benefited from great technological advances in detectors, optics, antennas, space instrumentation and computers. Since different physical processes produce radiation at widely different wavelengths, multi-wavelength and multi-messenger studies are essential to truly understand astronomical phenomena. Access to the wide array of telescopes available to Dutch astronomy is therefore a key element in its success.

Efficient use of resources is achieved by collaboration between the university astronomy departments (federated in NOVA), the institutes ASTRON for radio astronomy and SRON for space science, and the funding agency NWO, in particular the division of Physical Sciences (NWO-EW). These partners form the co-ordinating National Committee for Astronomy (NCA). Every decade the NCA formulates the national priorities in a forward-looking Strategic Plan. This Plan covers the decade 2011-2020, and shows how Dutch astronomy, within an essentially flat financial envelope, can maintain and even strengthen its excellent position in global astronomy.

To foster and keep talent, inspiring and rigorous teaching at the bachelor, master and PhD levels is needed in an environment with world-class facilities and a supportive funding climate. This requires a vital and sustained grants programme from NWO and the ERC, and a healthy funding of the universities.

Development of major astronomical instrumentation requires a long-term cycle spanning multiple decades and is driven by new scientific insights and technological possibilities. In this decade, the scientific harvest will focus on the high-priority instruments from the previous decadal plan: ALMA, VLT, LOFAR, Herschel/HIFI, Gaia and JWST/MIRI. Simultaneously, Dutch astronomers are playing an essential role in defining three key next generation international facilities, strategically chosen because they are the best match with available technical expertise and scientific interest so that involvement at PI level is both possible and natural.

These top priority facilities are ESO's giant 39-metre optical/infrared telescope E-ELT, the next-generation radio telescope SKA, and the far-infrared instrument SAFARI on JAXA/ESA's satellite SPICA: the astronomy facilities on the National Road Map Large-scale Research Infrastructure. Each facility provides strong scientific and technical opportunities for a prominent Dutch role. These are flagship projects which will require targeted investments. Structural direct funding of NOVA and funding through the NWO instrumentation programmes are essential for supporting facilities and enable smaller, shorter and more specialised projects, required for a vibrant community. Finally, to remain at the forefront in the decade after 2020, fundamentally new R&D concepts on which to base future instruments need to be developed now. Leadership in this long instrument development cycle is founded on membership of the international partnerships and treaty organisations ESO, ESA, ING and JIVE, and on secure, long-term funding for the research institutions charged with this work: ASTRON, SRON, NOVA and the university astronomy departments.

# 1 Scientific challenges in the 21st century

Astronomical exploration during the 20th century has shown the Universe to be a much more bewildering place than anyone had ever imagined. The Universe started its expansion in a hot Big Bang 13.7 billion years ago and has been cooling down ever since (Fig. 1). Almost 400,000 years after the Big Bang protons and electrons first combined to form hydrogen and helium and the Universe became transparent to radiation: the origin of the cosmic microwave background and the start of the 'Dark Ages'. Around 300 million years after the Big Bang the first stars and galaxies formed, lighting up the Universe and ionizing the intergalactic gas. All elements more massive than lithium, essential ingredients of life as we know it, have been synthesised in stars and supernovae. After multiple generations of stars, these elements ended up in the pre-Solar nebula out of which the Sun, the Earth and the other planets of our Solar system were formed by the slow accretion and coagulation of material in a cold disk of dust, gas and ices. Hundreds of planets around other stars have been found in the last two decades, and astronomers are closing in on a second Earth. The mass budget of the Universe is dominated by two mysterious components, dark matter and dark energy, that outweigh ordinary baryonic matter by 25:1. Unravelling their nature is one of the major goals of cosmology and fundamental physics.

The exploration and understanding of the Universe is far from complete. Using the contours sketched above, the challenge for the 21st century is to make sense of it all. Our current picture of the Universe evokes more strongly than ever those questions that are as much cultural as scientific: *How did it all start? Is the Universe infinite and ever-lasting? What are the fundamentals of space and time? How did quantum fluctuations grow to the structure of stars and galaxies we see around us? Where do we come from? Are we alone?* Observations in all windows of the electromagnetic spectrum and beyond, combined with the theory of the natural sciences and the power of modern computers and telecommunication, allow astronomers to address these questions in a scientific way.

Astronomy in the 21st century is a high-tech and computer-intensive science using state-of-the-art technologies, on the ground and in space. Astronomy is a driver of developments in faint signal sensor technologies as well as in major sensor arrays. It offers strong opportunities to precision technology industries. As such it links naturally to the economic top sector *High-Tech Systems and Materials.*

Dutch astronomy has a long standing tradition of planning and collaboration, reflected in a series of Strategic Plans, of which the current version covers the decade 2011-2020. The Strategic Plan is written on behalf of the National Committee for Astronomy (NCA), which is the informal coordination between the key players in Dutch astronomy: the universities, federated in NOVA, the NWO institutes ASTRON and SRON, and the NWO division of Physical Sciences (NWO-EW). It provides the underlying scientific landscape to the 'Toekomstplan Nederlandse Sterrenkunde 2011-2015', and extends the horizon to 2020. The science questions at the core of this strategic plan are also at the heart of the European ASTRONET road map.

In this Strategic Plan the Dutch astronomical landscape is described in Section 2, the key science questions are addressed in Section 3, required technology development in Section 4 and interdisciplinary links in Section 5. The importance of outreach and education is highlighted in Section 6. The main priorit-ies for the next decade are outlined in Section 7, with the consequent financial road map in Section 8. Sections 9 and 10 look back on the achievements of the previous Strategic Plan and look forward beyond the current horizon to identify expected major developments after 2020.

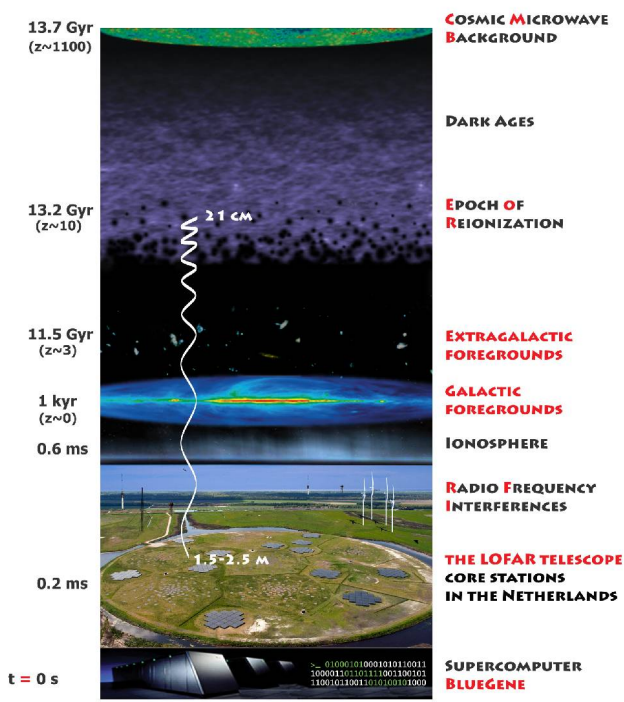

*Figure 1: A brief history of the Universe, starting 13.7 billion years ago just after the Big Bang at the Cosmic Microwave Background, followed by the first stars and galaxies at the Epoch of Reionisation 13.2 billion years ago, all the way to the current day when the EoR photons are to be detected with LOFAR antennas and processed in the BlueGene supercomputer in Groningen. Credit: Victor Jelic (LOFAR/ASTRON)*

## 2 Dutch astronomy: a reputation for excellence

In the global arena of astronomical research Dutch astronomy is regarded as of the highest quality, with a weight that significantly surpasses the size of the country and its astronomical community. This reputation is built on bright ideas in science and technology development; on a tradition of high-quality training of PhD students; on access to and a leading role in a wide scope of observational facilities across the electromagnetic spectrum, with a strong integration of technology development, instrument building and project management; on a keen interest of the Dutch general public in astronomy; and on a willingness of subsequent Dutch governments to strategically invest in Dutch astronomy when needed.

Particular indicators of the success of Dutch astronomy are the high number of prestigious US postdoctoral fellowships awarded to Dutch PhDs (both in a *per capita* sense as well as in absolute numbers, Fig. 2), the large number of Dutch astronomers that have received an Advanced Grant from the European Research Council (6 in the period 2008-2011, Fig. 2), the leading role of Dutch astronomers in global astronomical organisations, and the fact that Dutch astronomical institutes are consistently rated at the very highest levels by international review committees.

In 2010 the NOVA top research school, combining the university research groups, was evaluated as 'exemplary', the radio astronomy institute ASTRON and the space science institute SRON received the highest possible marks in their 2011 evaluations, and the 2012 evaluation of JIVE rated it as 'excellent'.

A key feature of Dutch astronomy is an integrated approach to the complete endeavour from idea to science result: from theory, scientific proposal, instrument and technology development and project management to student training, science harvesting and outreach. Although the Netherlands are a small country, Dutch astronomers frequently play leading roles in founding and running major observatories on the ground and in space: LOFAR, ESO, Herschel and ALMA are prime examples.

Maintaining this vanguard position is not a given, and requires a continued effort by all parties involved, particularly because instrument development cycle times are long (10-15 years) and are getting longer. Development of new technology, instruments and detectors is an essential way to influence the design parameters of new telescopes and gear them to the scientific needs of the Dutch community. This gives the Dutch astronomical community 'first pickings': the ability to reap immediately the highest profile science. Astronomical research is multi-wavelength- and multi-messenger-based, and maintaining access to a wide range of telescopes is therefore essential. Different physical processes radiate light at different wavelengths. To obtain a uniform and coherent understanding of astronomical phenomena access to all wavelengths is needed, from the very longest to the very shortest.

Astronomy is very much a global science and has a large appeal to the general public. This makes it a favourite to invest in for governments around the world, including those from the emerging economies (e.g., Brazil, South Africa, India, China). Global competition is therefore strong and growing. Building the largest telescopes requires international collaboration; at a European scale where possible and at a global scale where needed. At the same time, national initiatives can have a major impact. Two examples are LOFAR, a Dutch initiative in radio astronomy, now growing to global proportions, and the discovery of optical counterparts to gamma ray bursts which was made by a Dutch collaboration using the dedicated Italian-Dutch BeppoSAX space mission and the ING telescopes, run by the UK, the Netherlands and Spain.

### 2.1 Organisational structure of Dutch astronomy

To maintain the prominent Dutch role in astronomy, collaboration and coordination at a national level is a prerequisite. This informal coordination between the universities, ASTRON, SRON and NWO-EW takes place at the NCA.

*The Universities and NOVA:* The university research groups, at the University of Amsterdam, the University of Groningen, Leiden University and the Radboud University Nijmegen, have the dual mission of scientific research and student education, including the granting of PhD degrees. They collaborate in the top research school NOVA and coordinate their research agenda and instrumentation efforts through NOVA and the NCA. NOVA also functions as the national home base for the European Southern Observatory (home of the VLT, ALMA and the future E-ELT) and as such runs the optical-infrared instru-mentation group and the ALMA instrumentation group.

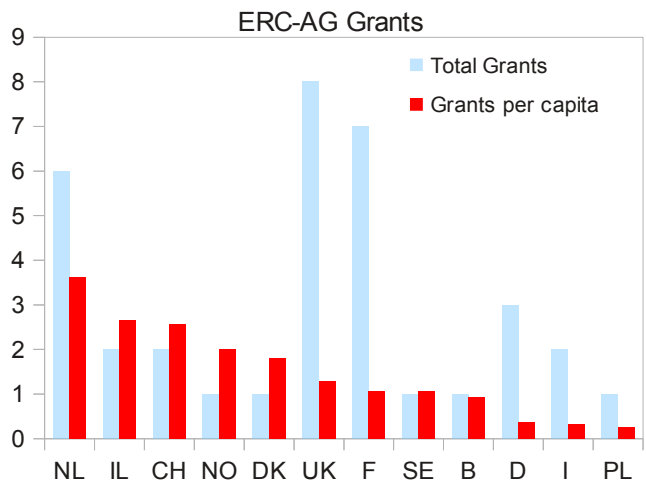 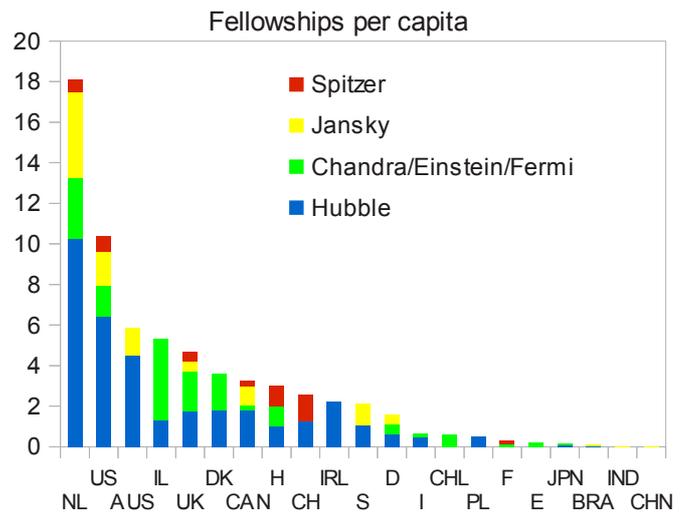

*Figure 2: Performance indicators of Dutch astronomy. Left, the total number of ERC Advanced Grants in astronomy per country in 2008-2011 (light blue), and normalised per inhabitant (red). Right, a similar plot of the total number of (highly competitive) US fellowships awarded by NASA to PhD graduates of different countries. In both cases the Netherlands can be seen to be leading the field, illustrating the high level and visibility of its astronomy programme.*

*NWO:* The NWO-institutes ASTRON (radio) and SRON (space) have a dual task of technology development, and scientific research, including operations support of observational facilities. Their technology and research programmes are coordinated with each other, and with NWO-EW and NOVA through the NCA. SRON is the national home base for the science programme of the European Space Agency (ESA). ASTRON hosts the Westerbork Synthesis Radio Telescope (WSRT) and the International LOFAR Telescope (ILT), and is the national home base for SKA activities.

The NWO division of Physical Sciences provides access to the Northern Hemisphere ground-based observatories, manages research and instrumentation grant competitions and interdisciplinary programmes, and funds the European collaboration in radio astronomy JIVE and the ALMA regional centre ALLEGRO. NWO-EW is also the national partner in bilateral agreements, in particular with emerging science nations (Brazil, South Africa, China, India).

A strong link between astronomical research, technical research and instrument development is essential for a vibrant community. Strong collaboration between, and even co-location of, technical institutes and research universities is highly desired to grow stronger bonds. Access of students across the country to active researchers and current instrumentation developments requires a balanced national distribution. Links with the Dutch technical universities are strong and growing further.

### 2.2 Funding of Dutch astronomy

Funding for Dutch astronomy is provided through three national channels:

– *Government.* The Ministry of Education, Culture and Science (OCW) is the national representative in the intergovernmental organisation for research in the Southern hemisphere ESO, the European Southern Observatory. The national contribution to ESO is therefore administered by the Ministry of OCW. The ministry directly funds the top research school NOVA as part of the *Bonus Incentives Scheme*. In addition the ministry provides the funding for the Dutch share of the science programme of the European Space Agency ESA (via the Dutch Space Office NSO). In recent years the government has also provided large competitive grants, in particular the BSIK funds which allowed for the construction of LOFAR and the National Road Map Large-scale Research Infrastructure, currently funding E-ELT and SPICA and with the SKA as an infrastructure for future funding. The Ministry of ELI (Economic Affairs, Agriculture and Innovation) funds technology development such as in the recently awarded DOME collaboration between ASTRON and IBM.

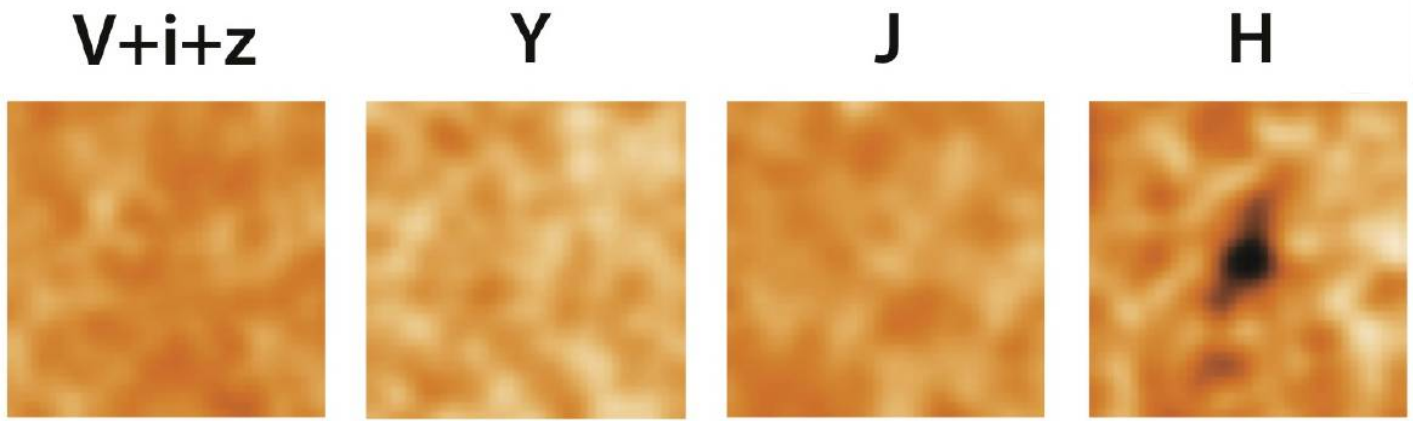

*Figure 3: The highest-redshift galaxy to date, in cut-outs from the Hubble Infrared Ultra-Deep Field at different wavelengths (V+i+z = optical, Y=1.1µm, J=1.3 µm, H=1.6 µm). The fact that the galaxy is detected in the H-band only, is a strong indicator that the source is at a redshift of z>10. Light that was originally emitted in the UV has stretched, due to the expansion of the Universe, to a factor 1+z larger wavelength, i.e. into the infrared. From Bouwens et al., 2011 Nature, 469, 504*

- *The Netherlands Organisation for Scientific Research (NWO).* NWO directly funds the national institutes ASTRON and SRON. The division of Physical Sciences funds the Dutch contributions to the ING observatory on the island of La Palma and the JCMT submillimetre telescope at Hawaii, to JIVE, and to the ALMA expertise centre ALLEGRO. NWO runs the proposal-based competitions where grants are awarded in competition with other sciences: larger personal grants in the Veni-Vidi-Vici scheme, individual projects in the free competition, and the funding schemes for instrumentation and technology development (NWO-Medium, -Large and the National Road Map Large-scale Research Infrastructure). NWO also awards the Spinoza prize, which has been won by six astronomers since the prize was introduced in 1995. Interdisciplinary research in astroparticle physics and astrochemistry is also funded through the NWO divisions of Chemical Sciences (CW) and Physics (FOM). Due to the increasing financial pressure on universities, PhD and postdoctoral funding now strongly depend on NWO, ERC and NOVA.

- *The Universities.* Astronomy is part of the curriculum at four universities in the Netherlands: the University of Amsterdam, the University of Groningen, Leiden University and the Radboud University Nijmegen. Through their collaboration in NOVA they coordinate the university research in astronomy and run the optical-infrared instrumentation group as part of NOVA's home base function for ESO. Many of the research staff at SRON and ASTRON hold a part-time or adjunct position at a university. Conversely, SRON and ASTRON partially fund staff members and PhD positions at the universities, leading to a larger integration of the national research effort.

Over the last years international funding has seen a strong growth. This has been particularly true after the start of the individual grant programme of the European Research Council and the success of Dutch astronomers in this programme: 6 Advanced Grants and 6 Starting Grants over the period 2008 – 2011. The EU Framework Programmes 6 and 7 are going strong, in particular for student training and technology development, and return funding to astronomy-related industry has increased from the European Space Agency and the European Southern Observatory.

## 3 Science questions for the next decade

The presence of the highest quality researchers, the highest quality training of the next generation of astronomers, and access to state-of-the-art instrumentation is, and always has been, the basis of excellence. Modern astronomical research requires a complete mix of theory, technology development, numerical simulations and multi-wavelength/multi-messenger observations. The Dutch astronomical community organises its research for the next decade around three major themes, each with a number of key questions, as outlined below. Strong links between the three themes exist, as well as with adjacent science disciplines: chemistry, high-energy physics and computer science. The choice of focus areas and key questions are a natural progression from those presented in previous Strategic Plans.

Significant evolution in the science questions continuously occurs as understanding grows and as new discoveries open new frontiers or shed new light on old issues.

A key asset of Dutch astronomy has always been its visible and well-connected position in the international community, making it one of the leading countries in astronomy worldwide. It is therefore not surprising that the research topics presented here are also at the heart of the European roadmap of ASTRONET for astronomy.

### 3.1 Galaxies: From first stars to the Milky Way

Tiny density fluctuations in the very early Universe, seen as temperature fluctuations in the cosmic microwave background, have grown into the rich complexity in structure (stars, galaxies, clusters) seen in the present-day Universe. How this structure formation has occurred is the overarching key question for the first theme. To answer this question requires a mix of theory, large-scale simulations and observations.

Major topics to be addressed are

– *First light, first stars: the epoch of reionisation*. When did the reionisation of the Universe progress and what were the sources of the ionizing radiation? How did the first stars and galaxies form and evolve (Fig. 3)? Can we detect gamma-ray bursts back to the epoch of reionisation and use them as tracers of massive star evolution in the early Universe? When did the first supermassive black holes form and can we detect them?

– *The formation and evolution of galaxies*. How do galaxies form and evolve? How did the first galaxies assemble themselves after the epoch of reionisation? What is the interplay between gas, radiation, stars, dust and supermassive black holes in an evolving galaxy (Fig. 4)?

– *The nature and distribution of dark energy and dark matter.* What is the nature and distribution of dark matter and what role does it play in galaxy formation and evolution? What is dark energy and when did it start to dominate the Universe? Are we heading for a Big Rip through the accelerated expansion of the Universe?

– *The structure and evolution of Local Group galaxies.* How did a galaxy such as our own Milky Way form? What is its structure and dark matter content? Can we infer the earliest phases of galaxy formation through galactic archaeology? How does it compare with galaxies in our Local Universe, and how can we understand the populations of stars in a galaxy as a whole?

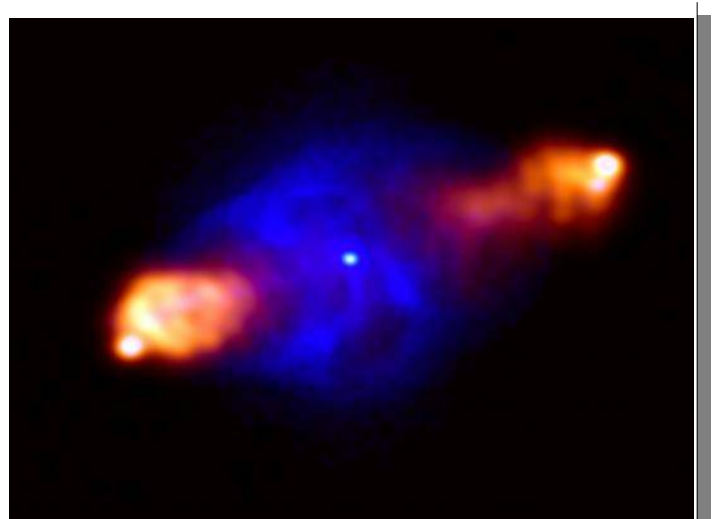

*Figure 4: Early LOFAR image of the radio jet in the active galaxy Cygnus A, observed at a frequency of 250 MHz. Credit: LOFAR/ASTRON*

The front line instruments now or soon to be online with which to address these questions are LOFAR for the detection of the hydrogen signal at the epoch of reionisation and the detection of the first black holes; the JWST space telescope for deep infrared imaging and spectroscopy of high-redshift and dusty galaxies and to trace the evolution of galaxies; the VLT and the ING telescopes for imaging and spectroscopy in the optical/near-infrared to understand evolving galaxies and their stellar populations; the VST for surveying the dark matter content of galaxies and galaxy clusters; and the WSRT-APERTIF for surveys of hydrogen in and around galaxies in the nearby Universe. Surveys for dust-obscured galaxies with Herschel and SCUBA-2 on the JCMT are transforming the field and are providing crucial sources for follow-up with ALMA. The ESA Gaia mission and large multi-object spectroscopy observations on 4-8m class telescopes will complement each other in mapping the galaxy distribution in the Universe, as well as characterise the stellar populations in the local Universe, in particular in our own Milky Way Galaxy.

In addition the coming decade will require investments that prepare for the next leap forwards: in the E-ELT to provide diffraction limited imaging and (multi-object) spectroscopy in the optical and infrared of distant galaxies as well as resolved stellar populations in the Local Universe; in SPICA for space-based far-infrared spectroscopy with which to trace star formation in the distant Universe; in the SKA to measure the signature of reionisation and elucidate its cause, and to trace the evolution of large-scale structure and of the hydrogen content of galaxies; in the Euclid space cosmology and structure formation mission for mapping the evolution of dark energy and the distribution of dark matter; in the ATHENA X-ray observatory for studying the nuclear black hole and galaxy cluster populations; and in the gravitational wave mission NGO to study the mergers and growth of supermassive black holes.

## 3.2 Towards life: stars, planets and their formation

Stars and planets form the natural habitat for potential life elsewhere in the Universe. Stars form in the cold, dense regions of interstellar space, when gas and dust clouds become unstable against the pull of gravity. A young star is surrounded by a rotating disk in which planets can form (Fig. 5). Over the last 15 years the fields of exoplanets and protoplanetary disks have exploded. Research is focused on theoretically and observationally understanding the full cycle of stellar life, planet formation and evolution, the formation of molecules in interstellar space and their incorporation into disks, leading up to detailed studies of exoplanets and their atmospheres.

Major topics to be addressed are

- *The formation of stars.* How do stars form out of the gas in galaxies? What triggers this process? What determines the distribution of stellar masses? Does this vary with location? How do young stars interact with their surroundings? How do the most massive stars form, how do their short lives progress?

- *Molecules: small and large.* How do simple molecules like water and surprisingly complex molecules like ethers and sugars form in the cold and tenuous clouds? What is the role of ice versus gas chemistry? Where do even more complex compounds like polycyclic aromatic hydrocarbons and buckyballs originate? How does the chemical composition of disks vary with amount of UV radiation and X-rays from the young star?

- *The formation of planets.* How do planets grow from tiny dust particles to pebbles, rocks and planetesimals and then on to full-blown planets? How does the dust composition evolve and when and where does crystallisation occur? What is the role of metallicity on planet formation? How and when do disks evolve from the gas-rich protoplanetary stage to the gas-poor debris phase, and thus lose their ability to form giant planets? What determines the architecture of planetary systems?

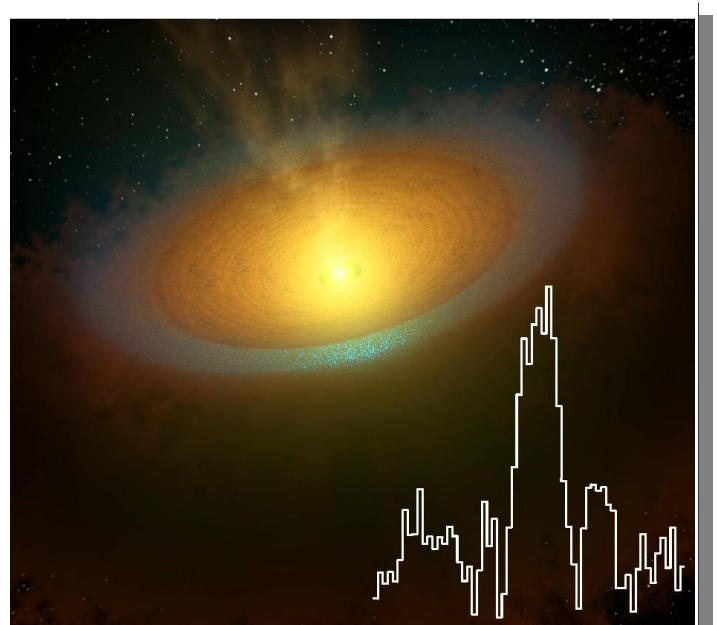

*Figure 5: Detection of the water reservoir in a forming planetary system, showing the $H_2O$ line at 557 GHz obtained with HIFI on the Herschel Space Observatory, detected in an annulus 100-200 AU wide in the disk around TW Hya (artist's impression in back). The amount of water is equivalent to thousands of terrestrial oceans. From Hogerheijde et al. 2011, Science, 334, 338) (credit:ESA/NASA/JPL-Caltech).*

- *The characterisation of exoplanets and their atmospheres.* What is the structure of exoplanets and how do they compare to the planets in our Solar system, in particular the Earth (Fig. 6)? Can we catch young planets in formation? What are conditions for life? What are the nearest planets to the Sun?

Current and near-future facilities that are providing a major step forward include the Herschel Space Observatory for an inventory of water and complex molecules in star-forming regions and searches for gas in disks, ALMA and JWST for zooming into the gas and dust distributions in the planet-forming zones of disks, and the VLT and ING telescopes for optical imaging and spectroscopy of young stars, disks and exoplanets, including the techniques of polarimetry and interferometry. A key element of these facilities is the combination of high-resolution spectroscopy with high-resolution imaging on solar-system scales. Laboratory experiments that provide basic spectroscopic information and that simulate the chemical processes under interstellar conditions are an integral part of the research.

Further major progress will require investments in the SPICA mission and the E-ELT to detect and survey gas and dust through far- and mid-infrared spectroscopy at unprecedented sensitivity and spatial scales;

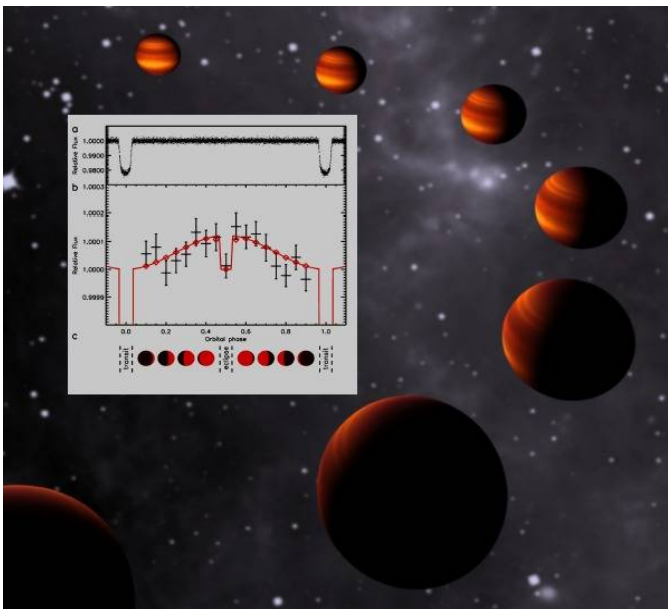

*Figure 6: The changing phases of extrasolar planet CoRoT-1b. Its night side hemisphere is consistent with being entirely black, just as seen for the interior planets in our own solar system. The wayside flux is most likely thermal emission (Snellen, de Mooij & Albrecht, 2009, Nature 459,543). Credit: de Mooij & Snellen.*

to directly image and characterise planets around other stars down to Earth-like candidates; and to obtain spectroscopy of the most massive stars in the Local Universe (Fig. 7). At centimetre wavelengths the SKA will be able to penetrate the dust in proto-planetary disks and image the inner regions of nearby systems. Gaia will chart the distances and movements of young stars in unprecedented detail. Complementary efforts in theory and numerical modelling remain crucially important. In the next decade, transit spectroscopy of exo-planetary atmospheres can be pushed to a next level using data from ECHO, a mission under study in the ESA M3 call. Transit and transient surveys reveal exoplanetary systems that orbit their parent stars in the plane of the sky and allow a precise characterisation of these planets.

### 3.3 The Universe as an extreme physics laboratory

The Universe is a place of extreme conditions that cannot be attained on Earth. It can therefore be used as a laboratory to probe new frontiers in physics, often in or near compact objects: white dwarfs, neutron stars and (supermassive) black holes. This is the environment where distortions of spacetime can be detected through X-ray timing and in gravitational waves, where cosmic-ray acceleration to extreme energies takes place in jets, and where the equation of state of supra-nuclear-density matter can be probed. Compact objects are the sites of the most energetic explosions known: supernovae, gamma-ray bursts and binary mergers. This also allows them to be used as probes for cosmology due to their enormous luminosity.

Major topics to be addressed are:

– *Fundamental properties of spacetime and matter.* Does general relativity hold in the strong-field regime around black holes and neutron stars? What is the equation of state of supra-nuclear-density matter as found in neutron stars? Can we use cosmic rays to understand particle physics at energies far exceeding those attainable in the LHC? Can we directly image the distortion of spacetime near the event horizon of a black hole? Can we detect and study the distortion of spacetime through gravitational waves (Fig. 8)?

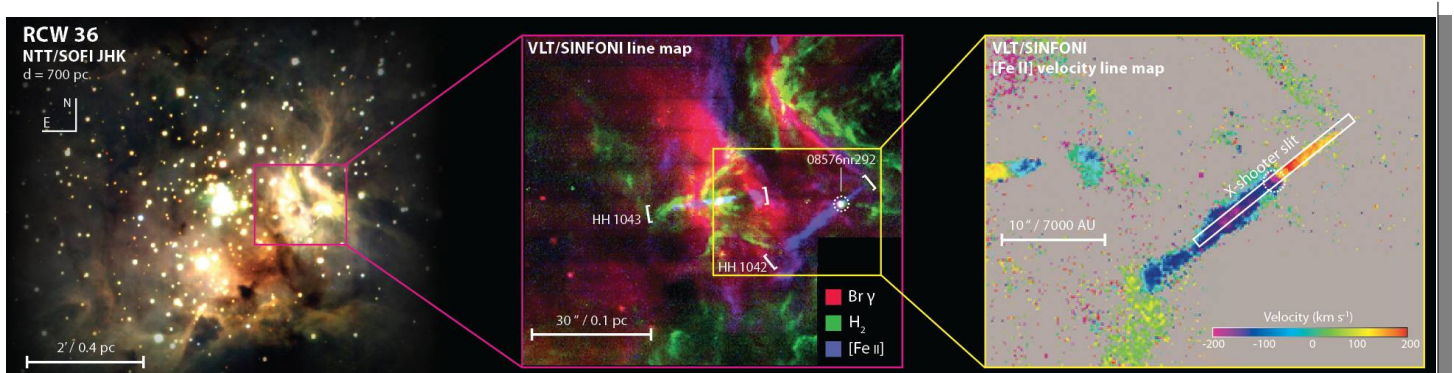

*Figure 7: Spectra obtained with X-shooter on the ESO Very Large Telescope of a massive young stellar object in the massive-star-forming region RCW 36 resulted in the discovery of jets, a signature that the object is still accreting material. Left: a near-infrared colour image of RCW 36; middle and right: a zoom-in on a VLT/SINFONI line map of 08576nr292 exhibiting two jets moving in opposite directions and producing [Fe II] emission. The large wavelength coverage of X-shooter allowed more than 300 emission lines produced by the disk/jet system to be analysed (from Ellerbroek, Kaper, Bik, et al. 2011, ApJ 732, L9).*

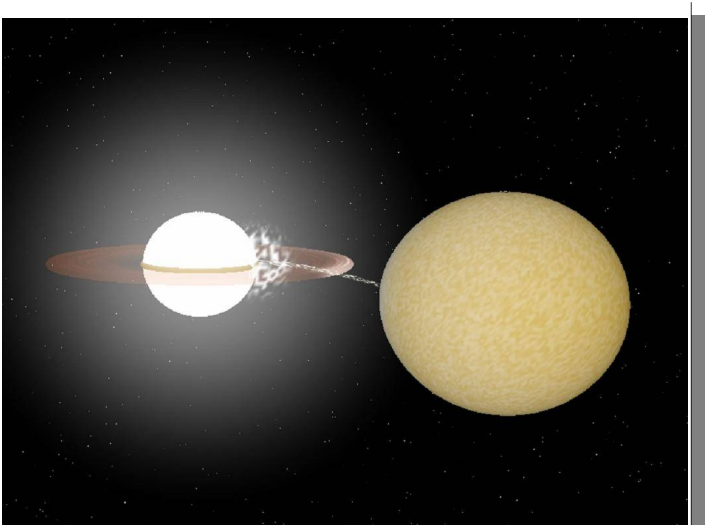

*Figure 8: Artist impression of the shortest period binary known, HM Cnc. In this system two white dwarfs orbit each other in a mere 5.4 minute period. The system will merge due to copious emission of gravitational waves in less than a million years. From Roelofs et al., 2010, ApJ 711, L138. Credit: Gijs Roelofs, Rob Hynes*

– *The physics of black holes and neutron stars.* How do black holes form? How are relativistic jets launched and how do they produce the highest-energy cosmic rays? What is the interaction of these jets with their surroundings? How do black holes merge and grow to masses of up to a billion times that of the Sun? What powers pulsars and magnetars?

– *The physics of accretion and accretion disks.* What is the physics of accretion? Can it be scaled up from binary stars to active galactic nuclei? Is the physics of accretion universal and independent of the nature of the central object? What is the nature of viscosity in accretion disks?

– *The transient Universe.* How does the evolution of stars lead to supernovae and gamma-ray bursts? In what type of binaries are white dwarfs triggered into a supernova type Ia explosion? How do neutron stars, black holes and white dwarfs merge and, possibly, explode? How often do these transient phenomena occur?

– *Stellar evolution and populations.* Which stars evolve into which remnant? How do the most massive stars evolve and end their lifes? What is the population of compact objects in a typical galaxy such as our own? How do they influence their surroundings? What are the source popula-tions of gravitational waves? How do stars form heavy elements essential for life?

Theoretical and numerical simulations are strongly pushing understanding of accretion flows and magnetic field evolution near neutron stars and black holes, complete stellar populations in the field and in globular clusters. Monte-Carlo simulations of air showers take into account radiative processes *ab initio* to trace observed cosmic-ray air shower properties to their primary particles. Observationally, current facilities include the X/γ-ray space missions XMM-Newton, INTEGRAL, Chandra, Swift, and Fermi for X-ray spectroscopy, timing and transient detection; the VLT and ING telescopes for optical/near-infrared identifications; LOFAR and the WSRT for cosmic-ray detections, pulsar observations and radio transients; ALMA for the study of supermassive black holes (Fig. 9); and Gaia for a characterisation of the stellar population in our Milky Way Galaxy. Strong astroparticle ties link astronomy with high-energy physics in LOFAR and the Pierre Auger Observatory for ultra high-energy cosmic-ray and neutrino research and with the VIRGO/LIGO communities for high-frequency gravitational wave detections.

In parallel the next steps will require investments in the SKA for the detection and characterisation of radio transients and pulsars as well as the study of supermassive black holes in the early Universe; in ATHENA, the next major X-ray mission for high resolution X-ray imaging and spectroscopy of plasma around compact objects; and in the E-ELT for resolved stellar populations of compact binaries and their progenitors in the nearby Universe. Using submillimetre VLBI the Event Horizon Telescope (EHT) will allow a direct imaging of spacetime distortions around the supermassive black hole in our own Milky Way Galaxy. Investments in the Čerenkov Telescope Array (CTA), NGO & VIRGO, transient surveys and LOFT present specialised opportunities for capitalizing on unique expertise in NL astronomy.

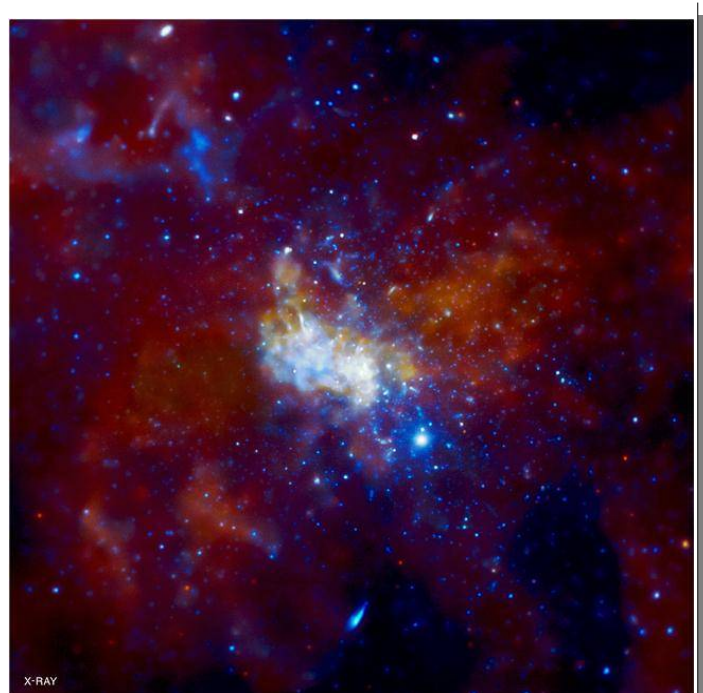

*Figure 9: Chandra observation of the centre of our Milky Way Galaxy, including the supermassive black hole Sgr A\*, which is seen to show small X-ray outbursts about once a day. Credit: NASA/CXO*

## 4 Astrophysics as a technology driver

Modern astronomy is a high-tech enterprise, both from the ground as well as in space. The newest technologies are developed and used in facilities that may look like a 'sensor array' as much as a classical telescope. System control technologies are as much part of a modern telescope as the best detector.

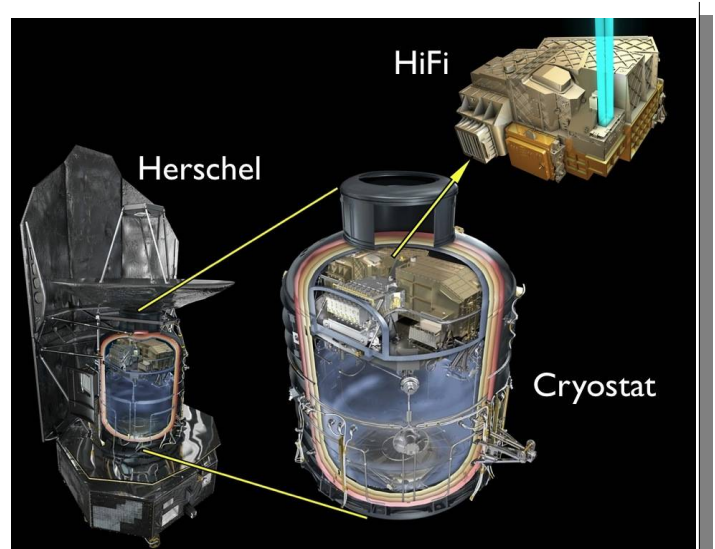

*Figure 10: Artist's impression of the HIFI instrument, built under the leadership of SRON, on board the Herschel Space Observatory. Credit: ESA*

Astronomical signals are almost always exceedingly faint, and detecting them against the strong background of natural and man-made emitters is a technological challenge. Bright ideas on the signatures of an astronomical event are to be combined with cutting-edge new technology. As such, astronomy is very much an idea- and technology-driven science as well as, at times, a technology-*driving* science.

Modern astronomy depends critically on advanced computing, and astronomers are among those pushing the envelope of ICT capabilities. Astronomers perform large-scale simulations with ever-more realistic physics and detail, and do so by using supercomputers around the globe, linked together in real time by lightening-fast internet connections. The ability to record, transport and analyse petabytes of observational data has led to wide-field surveys characterizing the large-scale distribution of galaxies in the Universe and the population of stars in our Milky Way Galaxy. The flow of data is such that some of the world's largest computer databases are astronomical in nature and astronomy is a favourite training ground for database engineering companies and supercomputer manufacturers.

Designing, building and operating new astronomical instruments requires frontier technical and computational solutions. Many of these solutions are the result of strong interactions between astronomy-driven concepts and industry-based quality assurance. Small high-tech industries invest in technological R&D for next generation astronomical instruments to better position themselves for new contracts with large companies like ASML and Philips.

In terms of the valorisation of fundamental research, it is worth to remember that the timescales for valorisation are usually long: decades, rather than years. A good example is the development of the WiFi IEEE 802.11 standard. At the root of this work is the development of new mathematical tools used originally for fundamental research: in particular, to facilitate detection of the radio signature of evaporating mini-black holes with the WSRT in 1977. The WiFi standard is now used in just about every wireless device in the market place, including laptops, mobile phones, satellite traffic navigators and digital audio broadcasting systems.

Astronomical techniques find spin-offs in other areas of science and business: adaptive optics techniques in eye surgery and radio-interference suppression in mobile phone communications are but two examples. Experience in e.g. the E-ELT ESFRI and ALMA/HIFI receiver programmes and in the Astrotec spin-off holding at ASTRON shows that in particular small and medium-sized enterprises (SMEs, in Dutch MKB) can benefit from fundamental research and technological developments in astronomy.

In the coming decade Dutch astronomy will foster and strengthen the connections to high-tech science in a number of areas, in particular:

– *Telecommunication.* The pioneering role of radio astronomy for wireless communication continues. Conversely, radio astronomy also uses the latest techniques in telecommunication for the detection

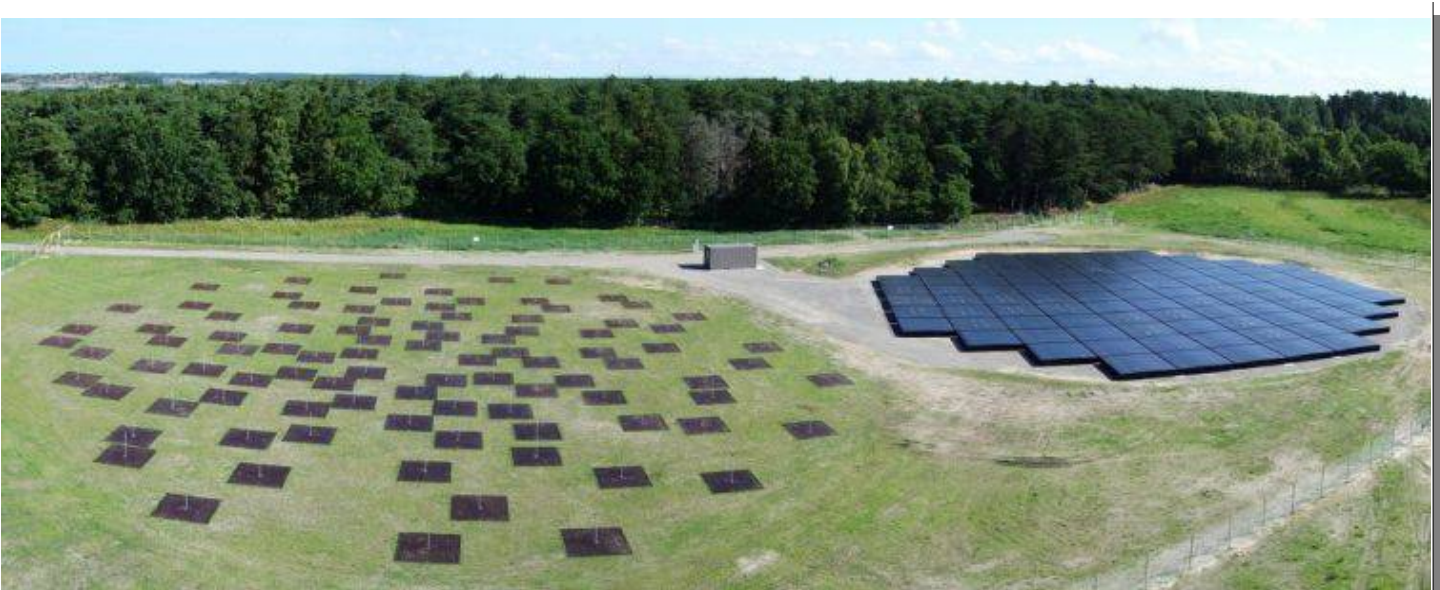

*Figure 11: The LOFAR station at Chalmers, Sweden: an example of how a modern telescope can take the form of a sensor array (brown, low-frequency tiles on the left; grey, high frequency pads on the right), and also how a Dutch initiative is spreading over all of Europe, linked by high-speed fibres for communication. Credit: ASTRON*

and transport of astronomical signals and the digital suppression of interference (e.g. the high speed network for LOFAR).

– *Information Technology.* How to effectively process and mine very large data flows has become very relevant over the last decade. Recently, the 33 M€, 5-year DOME project, a private-public initiative led by ASTRON and IBM, with support from the Northern Provinces and the Ministry of ELI was approved and will research the ICT aspects of the SKA Project in the Centre for Exascale Technology. The DOME project builds on the existing collaboration between ASTRON and IBM for LOFAR. Much ICT research activity on archiving and data mining is ongoing: an example is the TARGET programme in Groningen where multiple disciplines, academia and ICT industry are jointly developing massive data mining techniques, from which astronomy as a science can profit greatly.

– *Heterodyne detectors.* The Dutch community is world-renowned for the development and use of cooled heterodyne detectors, in particular in the far-infrared and sub-millimetre wavelength regimes. These are used on the largest and most advanced ground-based and space-based observatories, such as the ALMA array and the HIFI instrument on the Herschel Space Observatory (Fig. 10).

– *Cryogenic techniques and detectors.* Astronomers pursue ever higher accuracy, spatial and spectral resolution and thereby push detector sensitivities to their maximum. To suppress (thermal) noise many astronomical detectors and instruments need to be cooled down, sometimes to temperatures just above absolute zero. Fundamental development of cryogenic detectors is a long-standing effort, which opens the way for participation in the next major X-ray (ATHENA) and far-infrared (SPICA) space mission. In the optical-infrared regime particular attention is given to durability, light-weight and vibration-less systems, in particular through the E-ELT ESFRI collaboration with industry and technical universities.

– *Sensor arrays.* As the demand for larger apertures to detect weaker signals and to increase resolving power increases, individual astronomical telescopes are reaching a practical size limit. To further increase collecting area, telescope arrays must be used, coupled by high speed fibres for data transport. LOFAR is a prime example (Fig. 11), which will function as a prototype for the SKA, but also in the optical, mirror alignments and telescopes arrays are more common and required.

– *Precision engineering.* The high demands of astronomical instrumentation call for the highest-quality engineering. New techniques for the polishing of mirrors, alignments of surfaces, and manufacturing of highly-accurate movable elements need to be invented, tested and implemented. Examples are the aluminium mirrors developed for VLT/X-Shooter in NOVA's OIR Group and the slide mechanism in the Spanish EMIR instrument for Grantecan by Janssen Precision Engineering.

# 5 Interdisciplinary initiatives

The Universe is a place to be explored and, at the same time, to be used as a laboratory for understanding the natural sciences. Early in the 20th century this led to the highly successful cross-discipline of *astrophysics*. Now, new cross-disciplinary fields are emerging: astrochemistry, astroparticle physics, and, more into the future, astrobiology and exoplanetary atmospheres.

The Universe is, however, a peculiar laboratory. On the one hand we cannot influence experiments taking place in it, and, on the other, it is unique: there is no control lab. Cross-disciplines therefore always require a natural collaboration between astronomers and researchers from the other natural sciences.

Ever stronger links between astronomers and scientists in other natural sciences are growing, and to some extent are already established in Roadmaps of other disciplines, such as ASPERA (astroparticle physics) and EUROPLANET (planetary sciences). In the Netherlands collaborations are established or emerging, such as through the CAN (Committee for Astroparticle physics in the Netherlands), through NWO's astrochemistry programme, and through the new ALW-EW-SRON network on planetary and exoplanetary science. Investments in astronomy therefore benefit an increasingly wider scientific community.

Established connections are:

– *Astrochemistry.* The possibility for life elsewhere in the Universe is one of the biggest questions to mankind. The building blocks of life originate in outer space. New astronomical observations with instruments including Herschel and soon ALMA are uncovering solids, ices, water and ever larger organic (sometimes pre-biotic) molecules in the Universe. How do these species form and how far does chemical complexity go? How do they survive and evolve in the harsh environment of outer space? In turn, what do these molecules and solids tell us about the physical conditions in interstellar clouds such as temperature, density, and ionisation degree? State-of-the-art chemical physics experiments providing basic molecular data are essential to address these questions.

– *Astroparticle physics: cosmic rays, gravitational waves and neutrinos.* The 21st century faces the challenge of combining information from *multi-messenger* observations: photons, cosmic rays, gravitational waves and neutrinos. This joins high-energy physics with astronomy. Ultra-high-energy cosmic rays and neutrinos continuously bombard the Earth (Fig. 12). How do these interactions occur? Where do these particles come from? What do they tell us about the fundamental forces of nature? Multi-messenger studies of energetic sources such as active galactic nuclei and supernova remnants will reveal how these exotic objects produce cosmic rays. Gravitational waves are formed whenever mass experiences a changing quadrupole moment: in ultracompact binaries and stellar mergers, supernovae and gamma-ray bursts. Their detection is an enormous technological challenge, but in the coming decade the first events are expected. Opening up this window requires the combined forces of astronomers and gravitational wave physicists.

Emerging connections are:

– *Early-Universe cosmology*. The study of the origin of the Universe brings together theoretical physicists, high-energy physicists and astronomers. How did the Big Bang occur? What is the physics of the extreme conditions just after the Big Bang? How did inflation work? Where did the density fluctuations come from that form the basis of all struc-ture in the Universe? What is the role and nature of dark matter? Can we detect dark matter directly in laboratories on Earth?

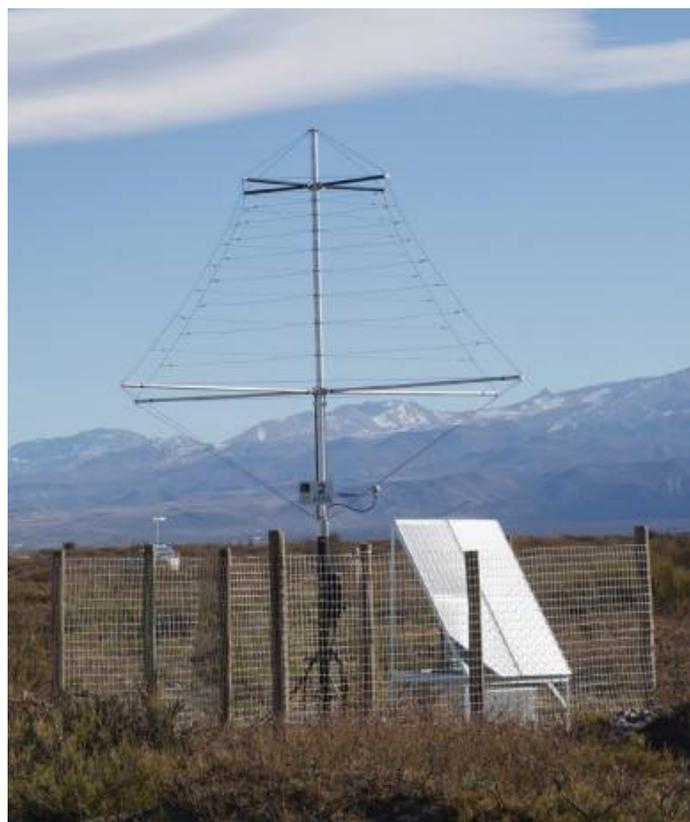

*Figure 12: An AERA station at the Pierre Auger Observatory, an example of a cross-discipline initiative to detect ultra-high-energy cosmic rays using radio astronomy techniques. Credit: IMAPP/RU Nijmegen*

– *Exoplanetary atmospheres and astrobiology.* The detection and characterisation of exoplanets offers the possibility of the new field of exoplanetary atmospheric physics and, perhaps, astrobiology. The characterisation of exoplanetary atmospheres requires the tools developed by atmospheric physicists to understand the atmospheres of the Earth and the other planets of the Solar system. Indeed, the rapid increase in our knowledge of the formation and evolution of planetary systems allows direct quantitative comparisons with the solar system, and cross-fertilisation between these fields will become more important. Do Earth-like exoplanets exist and how can we characterise them? Can we deduce the presence of life on these planets from the characteristics of their atmospheres? If life-tracers (e.g. ozone or elevated levels of oxygen due to photosynthesis) are detected, can we characterise and understand these life forms?

# 6 Education and Outreach

Astronomy directly combines modern-day high-tech science with a cultural curiosity into the beginning and end of everything, the start and future of life on Earth, and the uniqueness of humanity. Almost all children share a fascination for the Moon, the Sun, the stars and planets. In secondary education astronomy is a big draw for students into the natural sciences in general. The general public appreciates astronomy for its 'blue skies' aspects: it feeds our basic need for knowledge of the world and the Universe. Policy makers and industry see astronomy as a prime example of pure research without a direct application but with potential for technology spin-off. As a society we must remain willing to invest in excellent cultural and scientific endeavours, just for the sake of enriching our life. Excellence must be a guiding principle in these areas.

The Dutch astronomical community feels strongly the responsibility to inform and educate students and the general public about their work and the Universe. NOVA, ASTRON, SRON and NWO run outreach departments that are very effective in bringing astronomical news to the general public through websites, newspapers, television, social media and social gatherings. The NOVA Information Centre (NIC[1]) coordin-ates educational programmes for primary and secondary schools (Fig. 13) and offers outreach tailored to different audiences. Mostly these events are financed from base budgets and prize money (e.g. Spinoza prizes): these funds are not ear-marked specifically for outreach but 'freed up' because of the importance attached to it by the astronomical community. One prominent exception is the Universe Awareness programme, started in the Netherlands but now, thanks to subsidies from the Netherlands ministry OCW and the European Commission, an international programme that focuses on educating small children (especially from underprivileged communities) about our place in the Universe.

---
1 www.astronomie.nl

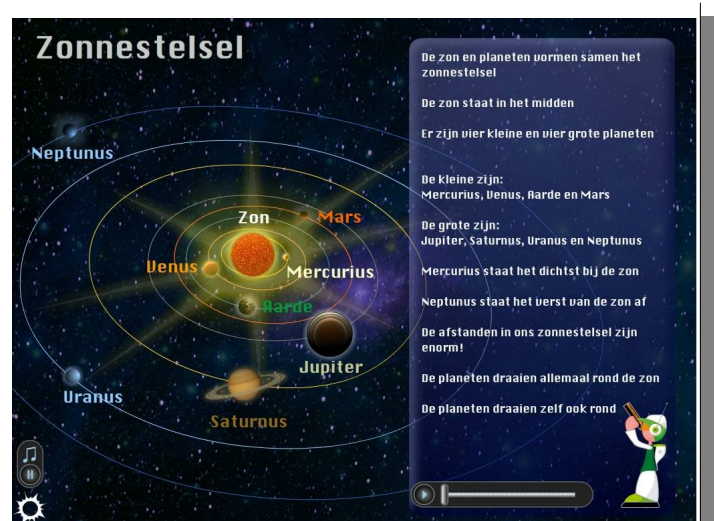

*Figure 13: Screenshot of the digiboard lesson on the Solar system, developed by the NOVA Information Centre for the highest grade classes in primary schools. An example of the outreach and educational efforts in Dutch astronomy. Credit: NOVA*

New high school physics curricula include attractive astronomy modules designed to stimulate interest in the natural sciences across the board. Since all natural sciences increasingly use the Universe as a laboratory, astronomy is a required ingredient of any university natural science programme.

At the Bachelor level the astronomy curriculum provides a broad overview of the workings of the Universe and its content. A priority for the next decade is to ensure that the existing Bachelor programmes in astronomy continue to receive full support, that an astronomy minor is offered to all natural science students, and that a more general introductory course in astronomy is made available to all university students.

At the Master level a specialisation in the programme is generally introduced, geared toward the specific re-search specialisation at a given university. A priority in the Master curriculum is to offer the students a co-ordinated and well-advertised programme at the national level, including educational opportunities at the NWO institutes and technical universities, as well as specialised programmes in the Management and Education tracks in the Master programmes.

An astronomy education develops capabilities much sought after by industry and business: analytical and solution-oriented thinking, helicopter views, priority sorting, and good mathematical and computer skills. Astronomy graduates can be found in many sectors outside (astronomical) research: government, management, consultancy, education and business, (Fig. 14).

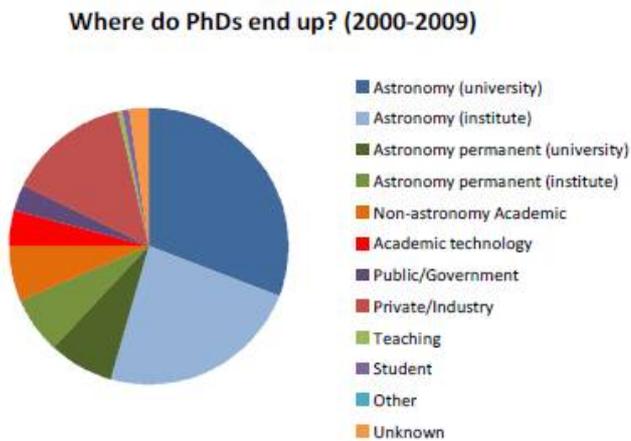

*Figure 14: Overview of current employment of 220 PhDs in Dutch astronomy between 2000-2009*

# 7 Priorities for Dutch astronomy in 2011-2020

To continue the world-leading role of Dutch astronomy and to accomplish the science goals as outlined in Section 3, an adequate infrastructure is needed. This infrastructure consists of a mix of scientists and students working at the universities and research institutes performing fundamental science and technology development, of engineers designing and building cutting-edge technology, and of software programmers who design and optimise the ever-more complex software needed to operate sensor arrays and store and analyse data and simulations.

Access to a wide array of wavelengths is an essential ingredient in the ability to answer the scientific questions that drive Dutch astronomy over the next decade. The required funding is a combination of the base funding for research, both fundamental astronomy as well as technology development at the NOVA, ASTRON and SRON institutes. A key component of a healthy infrastructure is therefore a stable and adequate baseline funding for these institutes. A major fraction of the astronomy PhDs as well as the Optical-Infrared Group, which is essential for the envisaged role in the E-ELT, are part of NOVA, which itself is guaranteed only until 2018. Long(er) term funding of NOVA is therefore a key priority.

Investments in manpower and talent are as essential as investments in facilities. Direct PhD student and postdoc funding has all but disappeared at the universities. Permanent staff are often temporarily on grants. NOVA and the grant programmes at NWO and at the ERC are the main channels for funding bright young talent in Dutch astronomy and they are therefore crucial to its success. At NWO this includes the Free Competition, the Veni-Vidi-Vici scheme and the TOP programmes. At the ERC these are the Starting Grants and Advanced Grants. The Netherlands has been extremely successful in these schemes, but they need to remain at least at the current level to ensure the scientific return on investment. Theoretical investigations and numerical simulations often form the basis of a scientific question or interpretation, but are never funded in big instrumentation projects and are therefore solely dependent on these channels.

Instrumentation development requires a long-term view, where science priorities need to be matched by technical expertise and project management skills. A number of major new facilities are currently in the planning phase, with first light foreseen early in the next decade. Matching the science questions with the technical and managerial expertise has led to three flagship projects where Dutch astronomy can claim a *leading, performance-determining role:* the E-ELT, the SKA, and the SPICA space mission (Fig. 15). All three offer a research infrastructure that serves a major part of the Dutch community, spanning the research themes described in Section 3. The Dutch community aims for a leading role in the instrument development for these facilities because it is the most effective way to influence the scientific capabilities of these facilities; because it offers large returns to astronomy related industry in the Netherlands; because it allows the Netherlands to remain close to the heart of international astronomy; and because it is fully in line with the technical development efforts taking place in NOVA, at ASTRON and at SRON.

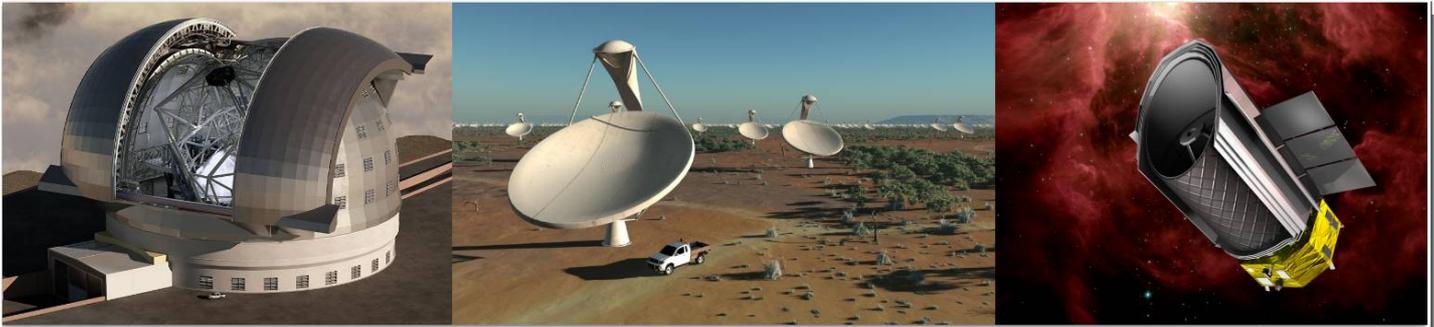

*Figure 15: The three astronomy projects on the National Road Map Large-scale Research Facilities. Left: the E-ELT on a Chilean mountain top above the clouds, middle: the SKA high-band dishes in a desert setting, right: the SPICA satellite in space. Credit: ESO/SKA/ESA-JAXA*

### 7.1 The E-ELT, SKA and SPICA-SAFARI

These three transformational projects are pivotal to the science questions of the next decades. The scale of these projects necessitates specialised investments. This is not a new situation: similar investments enabled the Dutch role in the VLT, Chandra, XMM-Newton, ISO, Herschel and LOFAR. In many ways these form the basis of our current success.

The flagship priorities are:

– **E-ELT:** The European Extremely Large Telescope is the flagship project of the European Southern Observatory. With a 39-metre diameter mirror it will be the largest optical-infrared telescope ever built and allows Europe to leap ahead of the rest of the world. Dutch astronomers aim to be PI on one of the E-ELT instruments (40% share) and to be a significant CoI on a second (20% share). The transformational character of the E-ELT lies in its collecting area and spatial resolving power in the infrared where it reaches its diffraction limit. Development of instrumentation for the E-ELT will be spearheaded by NOVA, and a major fraction of the investment needed for the instrumentation is already provided by the National ESFRI Roadmap grant awarded to NOVA in 2010 and the NOVA Instrumentation programme.

– **SKA:** The Square Kilometre Array is a globally-organised radio telescope which will ultimately combine a square kilometre of collecting area with a very broad frequency coverage, sited in Southern Africa and Australia. Dutch astronomers aim to lead the SKA development on phased arrays and focal plane arrays, a follow-through of the technology development set in motion for LOFAR and WSRT-APERTIF. The Netherlands also aims to be the lead country within the European collaboration for the SKA, using the JIVE collaboration as the seed for a European SKA operational headquarters. Within the Netherlands the SKA project is spearheaded by ASTRON, and the project will be phased over the next decade. Currently a major R&D development is ongoing, and from 2015 on the aim is for a 10% share in SKA Phase 1, which will be developed on the actual site. From 2020 on SKA Phase 1 will become operational, and preparation for its extension to the full array should be well-advanced.

– **SPICA-SAFARI:** The Japanese-European infrared SPICA mission will be a 3.2-metre diameter, actively-cooled telescope in space. Europe will provide the telescope assembly (ESA) and a consortium of European institutes will provide the SAFARI instrument which is based on technology developed at SRON over the last 15 years. SAFARI is an ultra-sensitive far-infrared imaging spectrograph making optimal use of the large collecting area of SPICA's cold mirror. The Netherlands are, with a 30% share, PI on the SAFARI instrument and lead the European collaboration.

The E-ELT, SKA and SPICA are all on the National Roadmap Large-scale Research Facilities.

### 7.2 Key contributions in a diverse environment

Much of the success of Dutch astronomy lies in its combination of theory, modelling, and observational access to all parts of the electromagnetic spectrum, now extending to astroparticles as well. In order to keep the research community vital, it is important that individual research and infrastructure grants continue to be available, outside the confines of large, long term projects. Risky, experimental projects and key niche contributions are an essential part of a vibrant scientific community. It is therefore imperative to maintain flexibility to join other projects without compromising the main objectives, as these may present

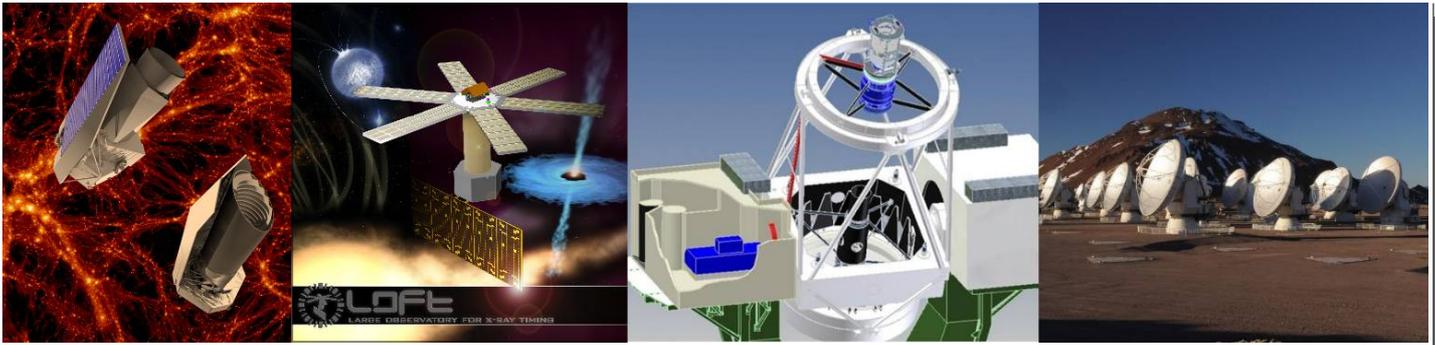

*Figure 16: A sample of projects to be financed through regular funding channels. From left to right: artists' impressions of the Euclid satellite, the X-ray timing mission LOFT, the multi-object spectrograph WEAVE on the WHT, and the ALMA array, which will function at the heart of the Event Horizon Telescope. Credit: ESA/ING/ESO*

unique scientific opportunities or be a test-bed for new technologies on which the next leap forward may be based. One such example is ATHENA for which long-standing detector development at SRON is a cornerstone element to the mission.

The timing of these missions and facilities is such that they cannot all be predicted at this time. As currently foreseen, strategically important projects where the Dutch participation can be funded through existing, regular funding schemes (such as NWO-Large/Medium) and which make efficient use of recent developments in technical capabilities and expertise include (Fig. 16):

– A leading role in the data processing and analysis in the **Euclid** cosmology survey mission, building on the OmegaCEN/AstroWISE development for surveys with OmegaCAM on the VST.

– Participation in a multiplexed **optical-infrared spectrograph** for the WHT/VISTA, VLT and/or the E-ELT, capitalizing on the expertise in the NOVA O/IR group on designing and building optical/near-infrared spectrographs.

– Participation in the next generation large X-ray facility **ATHENA**, under study as a candidate for an L-class mission in the ESA Cosmic Vision programme. SRON is leading an international consortium for the XMS instrument, which is an advanced X-ray imaging spectrometer using cryogenic detectors technology similar to those of SPICA-SAFARI.

– A leading contribution to the X-ray timing mission **LOFT**, currently under study as an M2-class mission in the ESA Cosmic Vision programme.

– Participation in **optical transient surveys**, e.g. PTF2 or the LSST, possibly coupled to electromagnetic-counterpart detections of gravitational wave sources.

– National initiatives in **interdisciplinary sciences** such as astroparticle physics (Pierre Auger Observatory, CTA, LIGO/VIRGO and NGO) and astrochemistry.

– The **Event Horizon Telescope (EHT)**, coupling a number of submillimetre telescopes, including ALMA, into a very-long baseline experiment to directly image the distortion of spacetime at the event horizon of the supermassive black hole in the centre of our Milky Way Galaxy.

– Fundamental **R&D** instrument development at all wavelengths to support large scale facilities, provide instrument for key, niche contributions and as entry-tickets into larger international collaborations.

### 7.3 Required support facilities

For the three flagship facilities (E-ELT, SKA and SPICA-SAFARI), as well as the facilities currently running or coming online in the next few years (VLT, LOFAR, ALMA, Gaia, Euclid, WSRT-APERTIF, JWST), structural support facilities and services are required to answer the science questions of Section 3. These include:

– Continued access and development of ground-based optical-infrared access to the Northern Hemisphere through the **ING** collaboration on La Palma. In particular follow-up capabilities to sources detected with LOFAR, Gaia and Euclid are essential. This requires a highly-multiplexed multi-object spectrograph on the WHT as well as longer term spectroscopic and photometric follow-up capabilities for individual sources.

– The **ALLEGRO** regional support centre for ALMA is currently entering its operational phase and will play a central role in allowing ALMA users to obtain the best science out of the ALMA array. If proven successful the ALLEGRO centre should be continued throughout the decade.

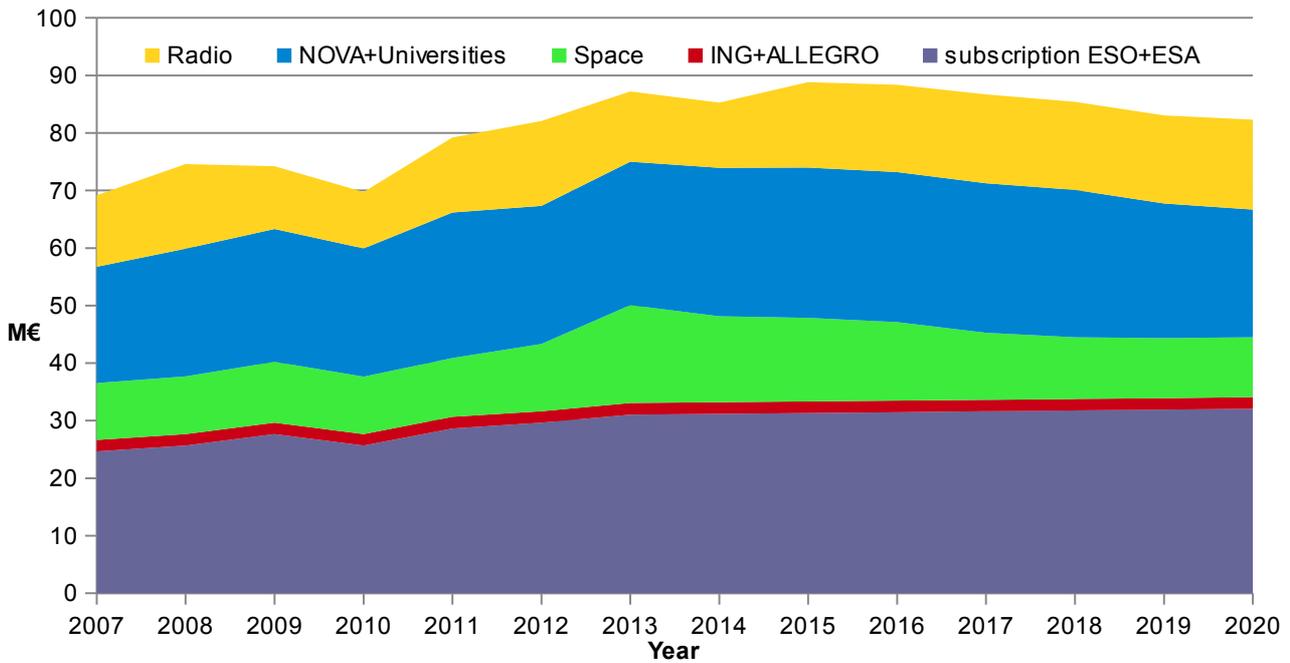

Figure 17: Total investment in astronomy in the Netherlands over the period 2007-2020. Numbers up to and including 2011 are actual expenditures. Numbers for the period 2012-2020 are forecasts in 2011 Euros, required to realise the plans presented here.

- The **JIVE** collaboration is the heart of European very-long-baseline interferometry at radio wavelengths, coupling radio telescopes all over Europe and as far as South Africa and China. Its expertise is not only an essential ingredient to the SKA and the EHT, but more generally provides the possibility of astronomy at unsurpassed spatial resolution.

*7.4 Ramp down of current facilities*

A number of mature facilities have reached the end of their scientific life time or are replaced by larger, more sensitive facilities. We therefore anticipate a withdrawal at the national level in the following facilities, although some may continue operations on a project basis by (inter)national collaborations:

- The James Clerk Maxwell Telescope on Hawai'i is scientifically overtaken by ALMA and Dutch activities will ramp down by 2013.

- The Isaac Newton Telescope on La Palma is expected to close down in the second half of the decade as a national facility.

- The Westerbork Synthesis Radio Telescope has been a premier radio observatory since the 1970s. After the current APERTIF large-scale surveys a ramp down is foreseen around 2020.

Furthermore, a number of international space missions, in which the Netherlands participate through ESA, are near the end of their life and will be discontinued before 2020. These include HST, XMM-Newton, INTEGRAL, Fermi, Chandra, and Herschel.

# 8 Financial road map

The science priorities outlined in Section 7 translate into a financial road map for Dutch astronomy for the coming decade.

Fig. 17 shows the total investments in base budgets at SRON, ASTRON and the university NOVA institutes, facilities, and exploitation required for realizing the ambitions in this Strategic Plan, and Fig. 18 shows the funding status. The budgets are in line with those given in the 'Toekomstplan Nederlandse Sterrenkunde 2011 – 2015'[2]. All budgets up to and including 2011 are actual budgets, not projections. The total over 2011-2020 is 848 M€, of which the majority is secured, either through base budgets or through already-allocated project financing. The fraction of the budget that is under consideration or remains to be secured is 14%. Note that the forecasted budgets are all in 2011 Euros and are not corrected for inflation. Fig. 17 shows a slight increase of 5% between 2011 and 2020, including two peaks due to investments in large-scale, global, infrastructure projects which are part of our top priorities.

---

2   Note that the ESA contribution was incorrectly mentioned in the Toekomstplan, which accounts for the difference of 4 M€ in the 2009 numbers.

Dutch astronomy has been very successful in leveraging the base budgets of SRON, ASTRON and the University institutes over the period 2005-2011, more than doubling them with regional, national and European grants and prizes for research and instrumentation. The same is true for the NOVA top research school, funded since 1998. The budgets presented here assume a continuation of this success. These external resources are, however, additive and can only materialise if the underlying foundation is sound and above a critical mass.

A number of components can be identified.

### 8.1 Base budgets and international contributions

– National contributions to the international treaty organisations ESO and ESA. The ESO contribution shows an increase from 2012, for ten years: while 60% of the E-ELT construction costs (1BEuro) will come from the regular ESO budget and from the contributions of new member states, 40% comes from the existing members in the form of a special contribution, with the Dutch share $1/20^{th}$ of that. The ESA contribution is flat after a slight increase in 2011.

– Base budgets of the university institutes, ASTRON and SRON are assumed to be flat. The ASTRON base budget also includes the costs of operating the LOFAR and Westerbork telescopes.

– National contribution to the ING, ALLEGRO and JIVE are all assumed to be flat for the coming decade. For the ING and JIVE incidental project budgets will be required as part of their instrumentation programme.

### 8.2 NOVA

– Funding for the NOVA top research school as part of the Bonus Incentives Scheme is assured until the end of 2018. The NOVA grant funds both research appointments and the O/IR Instrumentation group, and NOVA is responsible for the development of instruments for the E-ELT, ALMA and the VLT, as part of the national home base function of NOVA for ESO. Continued NOVA financing beyond 2018 is an important goal and it is shown as 'to be secured' in Fig. 18.

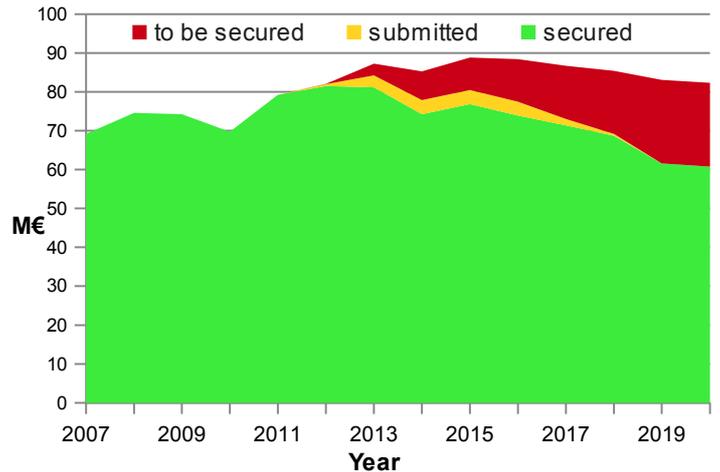

Figure 18: Funding status of the infrastructure requirement shown in Fig. 3. Budgets up to 2011 are actual expenditures. The 'to-be-secured' slice covers NWO-Large and National Road Map Large-scale Research Infrastructure grants that are still to be submitted/funded, and the current funding horizon of 2018 NOVA.

### 8.3 Flagship project costs

To realise our ambition with respect to the E-ELT, the SKA and SPICA-SAFARI, specialised project funding is required, outside of the regular NWO-based competitions (e.g. NWO-Large/Medium). We note that all three projects are on the National Roadmap for Large Research Infrastructure, with the timing of the projects such that funding requests are sequential and not parallel: E-ELT is largely funded, SPICA-SAFARI has been funded in early 2012, and SKA will be submitted for funding in the coming years.

– E-ELT: To realise the instrumentation programme for the E-ELT a total of 35 M€ will be required. The finances for this NOVA-led programme are secured up to 2018, with the exception of instrument hardware costs (2 M€ to be provided by ESO and 2 M€ from NWO grants) and NOVA financing beyond 2018. The uncertain future of NOVA after 2018 represents 6 M€ which are currently unsecured.

– SPICA-SAFARI: The Dutch contribution to the SAFARI instrument on the SPICA mission will amount to a projected cost of 60 M€. Funding for SAFARI mostly comes from the SRON base budget (36 M€), and 18 M€ has been secured within the National Roadmap for Large Research Infrastructure 2012. The remainder will either have to be cut from the project costs, or be requested at a later stage. The full cost is included in Figs. 17 and 18.

- SKA: The SKA project in the Netherlands can be divided in three phases: SKA R&D (currently running), SKA1 (2015-2019), and SKA2 (2020+). Funding for SKA R&D is to be provided through the Collaboration of the Northern Provinces (Samenwerking Noord Nederland) and an NWO-Large application (6M€). National Roadmap for Large Research Infrastructure funding is required for SKA1 (30.5M€). SKA-2 is expected to have an operational budget of 3 M€/yr, but falls beyond the horizon of the current plan. The private-public DOME project for the SKA has been financed in early 2012, and has not been included in the figures here.

### 8.4 Instrumentation through regular funding schemes

Regular funding lines for instrumentation mostly includes the NWO competitions of NWO-Medium and NWO-Large. The instrumentation programme within NOVA offers a time-limited possibility for instrument development.

- NWO-Medium: NWO-M has a yearly round with a total budget of 1.6 M€/yr. It is *the* prime funding line to maintain the diversity of the research environment and smaller-scale niche contributions to international projects. NWO-M is an NWO-EW funding line, with highest pressure from astronomy, and is expected to remain at the 1.5 M€/yr level (Fig. 19). A recent inventory showed an expected funding pressure of 38 M€ for the next 5 years: an oversubscription factor of 5 if NWO-M is kept at current levels.

- NWO-Large: This NWO grant line for all areas of scientific research runs every two year for projects above 1.5 M€. The total available budget is about 20 M€ per call. The plans presented here represent one astronomy project per call with a budget of between 3-7 M€ per project.

- NOVA-4 Instrumentation: The major part of the NOVA Phase 4 instrumentation budget (2013-2018, 11 M€) will be used for E-ELT instrumentation and the Euclid ground segment (7 M€ in total). The remaining funds will be used for strategic investments for key contributions and seed funding, aimed at stimulating the instrumentation expertise at the universities. The O/IR Laboratory will also perform a number of R&D projects and, jointly with NWO, will work for the ING (2 M€ NWO-EW).

- ASTRON, SRON and NOVA are part of the Topsector *High-Tech Systems and Materials*, which will offer opportunities, in particular for R&D and spin-off activities.

### 8.5 Research grant competitions and prizes

With ever tighter university budgets the financing of research (theoretical, numerical and observational) depends more and more on grant competitions and prizes. The financial roadmap is based on continued success in acquiring funding from such channels.

- NWO Free Competition: The NWO-based Free Competition is currently the only grant programme that is open to all researchers, without restrictions, and is therefore essential for a diverse research environment. Through the Free Competition astronomy currently receives 1.2 M€/yr, which funds 6 PhD or postdoctoral positions: an average of one such position per 10 years per staff member, with an oversubscription rate around 5. The total budget is below the minimum required and not even all projects rated 'excellent' can be funded. An effective programme should be double the size with a funding percentage near 20%, such that the excellent projects can be funded.

The TOP programme was introduced by the division of Physical Sciences, but due to budgetary constraints a call has not been issued yet. TOP grants fund more than a single position such as in the Free Competition, and is expected to be 750 k€ per grant. When funded properly, astronomy expects to be able to win grants at a level of 2 grants per call, which may alleviate the pressure on the Free Competition, if applications to the TOP grant

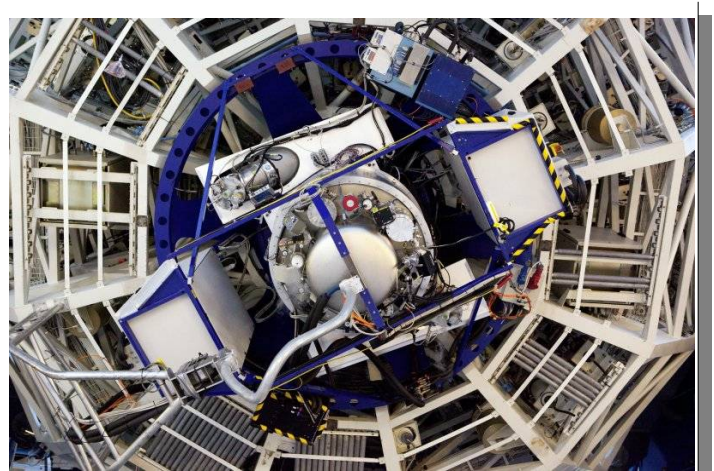

*Figure 19: The successful X-Shooter spectrograph behind the spiderweb-like mirror cell of the VLT at ESO Paranal. The Dutch contribution to X-Shooter was financed through NWO-M and the NOVA instrumentation programme. Credit: NOVA*

are unrestricted with respect to science area, career status and age.

- NWO Innovation Scheme: The Innovation Scheme (Veni, Vidi, Vici) has been very successfully applied to astronomy and is a major factor in the continuing rejuvenation of Dutch astronomy as it provides a strong impulse for new faculty, in particular through the Vidi scheme. Recent Vidi recipients are now moving into the Vici regime and pressure on the Vici grants (typically 1-2 per year for astronomy) will increase, certainly in the first five years of the coming decade. Veni grants are currently not as effective as they could be due to the awkward timing of the Veni deadlines with respect to the global deadlines for postdoc positions, which close in November with grant decisions in early February. A level of 1-2 Vici, 3 Vidi and 4 Veni grants per year (4.5-6 M€/yr) for astronomy will be required for a continued stimulus of young and new staff.

- ERC Grants: Starting, Advanced and Synergy: The ERC Starting and Advanced Grants are an increasingly important funding line, where Dutch astronomy is doing exceptionally well (see Fig. 2). This funding is used both for research and instrumentation and is a funding possibility for the projects mentioned in section 7.2. Extrapolating the current track record, 1 Advanced Grant and 1 Starting Grant a year should be sustainable (4 M€/yr). Further initiatives from the ERC, such as the new Synergy Grant scheme (started in 2012), provide opportunities for larger projects, in particular in interdisciplinary fields.

- Spinoza Awards: The Spinoza awards (2.5 M€) are the most prestigious science awards in the Netherlands and not a grant competition. Five Dutch astronomers have won the award since its introduction in 1995. Spinoza awards are used for a mix of research, instrumentation development and outreach (Fig. 20).

### 8.6 Funding requirements by agency

These funding lines translate into the following requirements, grouped by agency/organisation.

*Government (OCW)*

- Annual contribution to treaty organisations ESO (9 M€/yr for the period 2012-2022) and the astronomical science budget of ESA (23 M€/yr). The bulk of this funding returns to Dutch industry as contracts for building telescopes, satellites and infrastructure.

- Structural funding or continuation of NOVA beyond 2018 (5 M€/yr), both for its coherent national research programme as well as for its optical-infrared instrumentation programme, an essential part of NOVA's role as the national home base for ESO.

- Incidental funding opportunities for very-large research infrastructures (E-ELT instrumentation, SKA and SPICA-SAFARI) that cannot be funded through regular grant schemes. This includes programmes awarded (E-ELT, SPICA-SAFARI) and to be funded (SKA) within the National Roadmap for Large Research Infrastructure.

*NWO*

- Continued base budget funding of SRON and ASTRON at their current levels with inflation correction.

- Continuation of the Veni-Vidi-Vici Scheme or an equivalent programme at adequate levels and with deadlines and evaluation periods in line with global standards.

- Continuation of NWO-Large at the current levels and at the current rate.

*NWO-EW*

- Doubling of the number of astronomy research positions funded in the annual Free Competition to 12 fte/yr (a goal from the previous two plans that is still not attained).

- Structural continuation of the NWO-Medium investment subsidies at a level of 1.5 M€/yr.

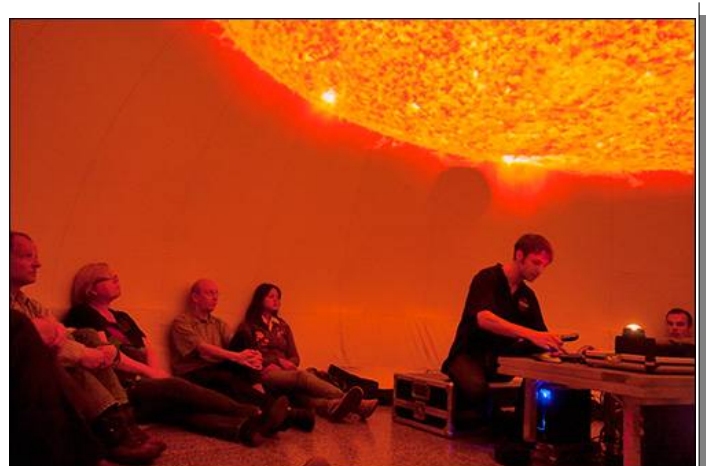

Figure 20: The NOVA mobile planetarium which has been visited by more than 30000 school children already, partially made possible through Spinoza awards. Credit: Dick van Aalst/RU

- Grants of Veni-Vidi-Vici positions to astronomy at a rate commensurate with the number of excellent researchers in the field (approximately 1-2 Vici, 3 Vidi, 4 Veni per year). Annual competitions for TOP and rolling grant subsidies at a level of 1.5 M€/yr for astronomy.
- Continuation of ING, ALLEGRO and JIVE at current levels, with incidental grants for new instrumentation.
- Stimulate cross-disciplinary research by reserving funds at 1.5 M€/yr to continue or start challenging programmes in astroparticle physics, astrochemistry, (exo)planetary atmospheres, and computer science.

*Universities*

- Maintain research staff at least at the current levels to enable the scientific harvesting of current results, to train the next generation of astronomers, and to optimally position Dutch astronomy to remain at the forefront world wide.

*ERC, Marie Curie & Erasmus Mundus*

- Continuation of the ERC programmes at their current levels and in the current open and flexible way.
- Continuation of the Marie Curie and Erasmus Mundus programmes which provide crucial opportunities for mobility and training of young students and researchers across Europe and associated member states.

The combined and coherent funding by the above-mentioned agencies and institutions will allow Dutch astronomy research to thrive, new technologies to be developed and implemented, and will offer a long-term sustainable future for Dutch astronomy at the very forefront of global science.

# 9 A retrospective: 2001-2010

The previous Strategic Plan of Dutch astronomy covered the period 2001-2010. In the following we briefly review its main priorities, followed by their current status:

*A strong collaborative research structure between NOVA, ASTRON and SRON.*

*Status: mostly achieved.* Both ASTRON and SRON have linked their research departments to the university groups via cross-appointments and new hires at PhD and staff level.

*Continuation of the NOVA top research school.*

*Status: achieved.* NOVA was evaluated and found to be of the highest quality in 2005 and again in 2010. Funding for NOVA is currently secured until 2018, with continuation as long as NOVA remains *exemplary*, as specified in the policy plan 'Kwaliteit in verscheidenheid' of the ministry of OCW in summer 2011.

*Participation in ALMA through ESO*

*Status: achieved.* ESO committed to building ALMA in 2002, in collaboration with the US and East Asia. The Netherlands has successfully concluded the design, construction and delivery of 64 Band 9 receivers to the ALMA observatory, led by NOVA and SRON (Fig. 21). ALMA has been under a Dutch director since 2008 and started first scientific operations in the second half of 2011.

*ESO VLT/VLTI instrumentation*

*Status: achieved.* The VISIR, MIDI, X-Shooter (Fig. 19) and Sphere instruments have been delivered for the VLT(I), and construction of MATISSE and MUSE is under way. These activities are now under the umbrella of NOVA after the reorientation of ASTRON. Since NOVA is not structurally funded this is a precarious situation that remains to be resolved. The

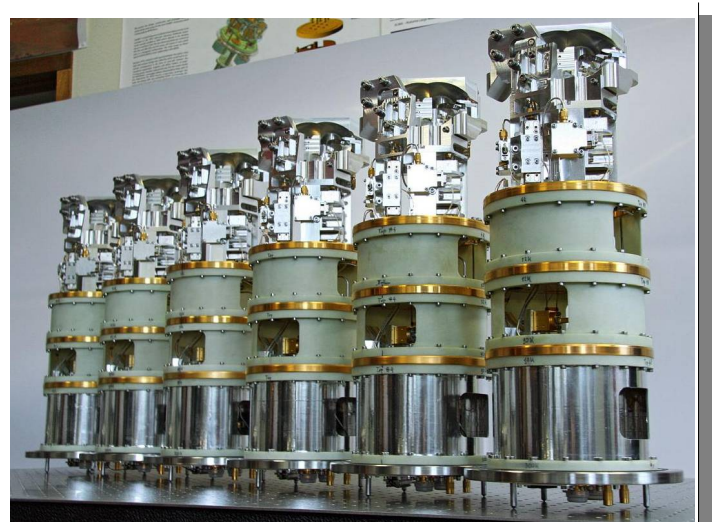

*Figure 21: Six of the 64 ALMA Band 9 cartridges, a main part of the Dutch contribution to the ALMA project, designed and produced in a collaboration between NOVA and SRON. Credit: NOVA/SRON*

OmegaCAM camera for the VST was developed under Dutch leadership and delivered to ESO.

*NGST (=JWST) mid-infrared camera/spectrograph*

*Status: achieved.* The cold bench of the spectrograph on the MIRI instrument was designed and built under Dutch leadership and successfully delivered to the JWST consortium (Fig. 22). It awaits the JWST launch.

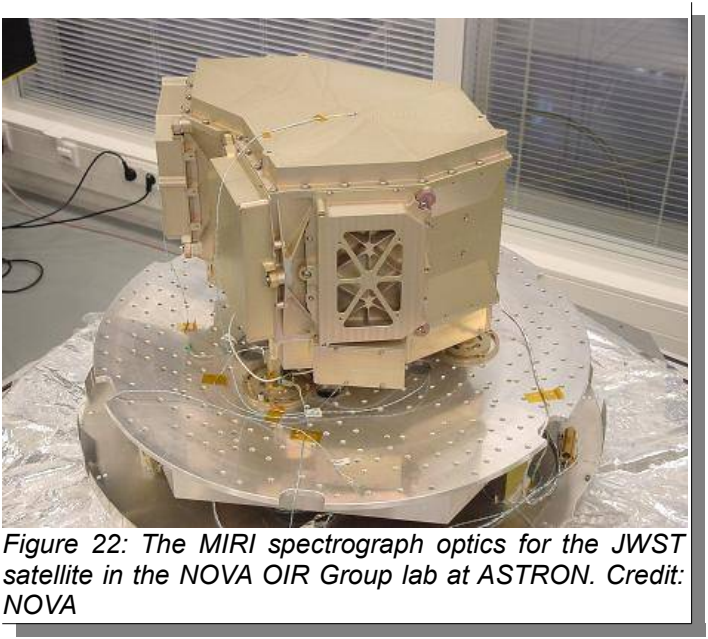

*Figure 22: The MIRI spectrograph optics for the JWST satellite in the NOVA OIR Group lab at ASTRON. Credit: NOVA*

*LOFAR/SKA preparation*

*Status: achieved.* LOFAR hardware design and development was funded through a BSIK grant from the Dutch government and the BlueGene supercomputer was (partially) sponsored by IBM. Software development for its astronomical use has been achieved in combined effort by the universities, ASTRON, NOVA, NWO-Medium and NWO-VICI and ERC grants. LOFAR started operations in 2011 and is developing into the International LOFAR telescope (ILT), including Germany, Sweden, the UK and France. The SKA has seen major developments over the last few years, in particular, the establishment of the SKA Organisation Ltd that is funded by national funding agencies (including NWO and the ministries ELI and OCW) and is charged with bringing Phase 1 of the project to construction readiness in 2015.

*JIVE real time VLBI*

*Status:achieved*. VLBI real time measurements were achieved using the a global e-VLBI array in 2007 and are now the normal mode of operation for JIVE.

In addition, a number of other opportunities that presented themselves after the 2000-2010 plan was written were also realised. SRON was PI on the HIFI instrument for the Herschel Space Observatory, the most complex space instrument developed so far in the Netherlands. HIFI was successfully finished and launched and is performing to the highest standards (Fig. 10).

In the first phases of the E-ELT project definition the Netherlands successfully participated in four Phase A instrument studies, laying an exciting foundation for future involvement. In 2009 NOVA won a government/NWO investment grant for R&D and instrument development for the ESO E-ELT as part of the ESFRI initiative. In 2011 ESO selected the METIS mid-infrared instrument as one of the first three E-ELT instruments, with a Dutch PI role, and the Micado near-infrared camera as a first light instrument, with a Dutch Co-PI role (Fig. 23). A Dutch participation in an optical-near-infrared multi-object spectrograph and an exo-planetary imaging instrument remain as options, and could be the subject of funding proposals in existing programmes (e.g., NWO-M/L) later in the decade.

Another unforeseen and positive change in the past decade was the reintroduction of the astronomy programme at the Radboud University Nijmegen in 2001, which joined NOVA in 2003. In contrast, 2011 saw the closure of the astronomy department at Utrecht University, which led to a redistribution of most of its staff over the remaining NOVA institutes, but represents a net loss of national research capacity.

Developments foreseen in the 2000 Strategic Plan that did not blossom in the Netherlands in the way anticipated are computational astrophysics as a separate field, adaptive-optics developments and an increase in the NWO funding for PhDs and postdocs.

Computational astrophysics did not develop into a separate field, but has become an almost indispensable part of astrophysics in general. Development of computing power, communication speeds and storage capacity outpaced its anticipated speed, delivering much more computing resources to the average user than expected. Computational astrophysics is therefore a more commonplace practice than just a niche, and in fact it is now a part of the astronomy MSc curriculum. Nevertheless, high-performance simulations of physical phenomena in the Universe, astronomical databases, and correlators for ever-larger interferometers keep pushing the envelope of even today's immense capabilities.

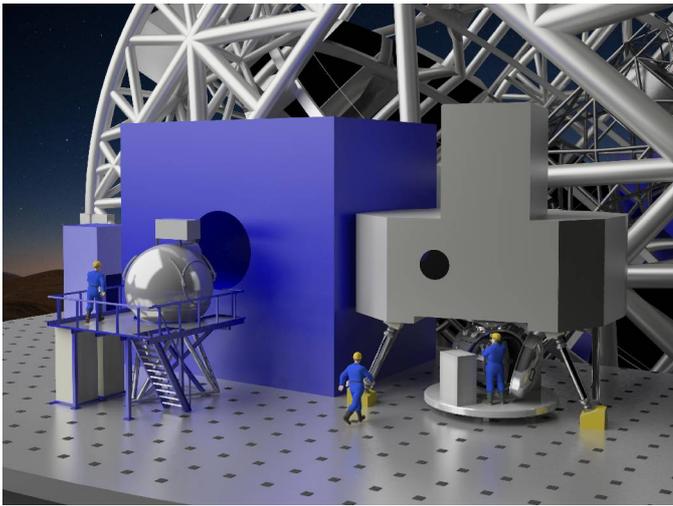

*Figure 23: Artist's impression of the proposed Metis instrument (sphere at left) and the Micado imager (near man on right) at the Nasmyth platform of the E-ELT telescope. Credit: NOVA*

The adaptive-optics efforts in the Netherlands remained relatively small since available resources were not sufficient to catch up with long standing efforts abroad.

NWO free competition funding for PhDs and postdocs is tighter than ever, and this is a worrisome situation. On average only 6 positions per year are available for all of Dutch astronomy. The impact of this sub-minimal funding has been somewhat mitigated by very successful applications for large personal NWO and ERC grants. There is a risk that the resulting increasingly-uneven distribution of funding will lead to a desertification of the community.

## 10 A forward look: 2020+

Time scales for major international astronomical projects are now often longer than a decade. Developments for longer term projects are therefore already under way, and a forward look is required. The three main priorities for funding in the current decade (E-ELT, SKA and SPICA) will see first light around or after 2020, and the science harvesting with these facilities, is therefore the topic of the next Strategic Plan.

Following the developments of the research themes in Dutch astronomy, a number of key facilities on the drawing board for after 2020 can be identified (Fig. 24). They are approximately listed in the chronological order in which they may appear.

In the era of the E-ELT the current flagship facilities of ESO, the VLT and ALMA, are still required and will remain highly productive. It can therefore be expected that a call for instruments for $3^{rd}$ generation VLT instruments and $2^{nd}$ generation ALMA instruments will be issued. In line with the successful Dutch instrumentation programme on both facilities a leading role in a (multiplexed) optical-infrared spectrograph for the VLT and a high-frequency instrument for ALMA can be expected. The E-ELT itself can be equipped with a total of 8 instruments at two focal stations. The current instrumentation plan calls for instruments 1-5, and among the remainder is an exoplanetary imaging instrument, which would be a natural follow-up on the NL-led Sphere-ZIMPOL instrument for the VLT. A high-efficiency natural seeing optical spectrograph would also be a natural complement to the instrumentation suite.

Synoptic survey astronomy may enter a new era with the LSST, and arrays for almost continuous all-sky monitoring of the optical/infrared sky will provide opportunities for faint transients and massive statistical studies of galaxies: expertise in data mining is essential to exploit these possibilities, as well as appropriate (spectroscopic) follow-up facilities.

Gravitational-wave astrophysics is expected to be a major growth area in the next decade. After a likely successful detection of gravitational waves with LIGO/VIRGO this decade, the NGO satellite and the ground-based Einstein Telescope will be natural successors. These facilities not only require major (industrial) investments, but also supporting astronomical facilities in radio, optical and/or X-ray wavelengths, fitting very well in a cross-discipline collaboration between Dutch physicists and astronomers. SKA will enable study of the longest- wavelength gravitational waves through pulsar timing.

In X-rays, next generation space missions such as ATHENA and LOFT will be a major step forward in sensitivity, spectral resolution and timing, further pushing the studies of compact objects, active galactic nuclei and the hot gas in galaxy clusters. The Čerenkov Telescope Array will revolutionise photon astronomy at TeV energies, revealing extremely high-energy phenomena around pulsars, magnetars, supernova remnants and hypernovae.

The hunt for a second Earth will only intensify in the next decade through transit spectroscopy and direct imaging space missions will be launched in the 2020+ time frame. Currently foreseen missions include the Exoplanet Characterisation Observatory (EChO), in the running as an ESA M3 mission, and more advanced, space-based interferometry missions can be expected. Both would fit very well within the Dutch developments on exoplanet imaging and characterisation.

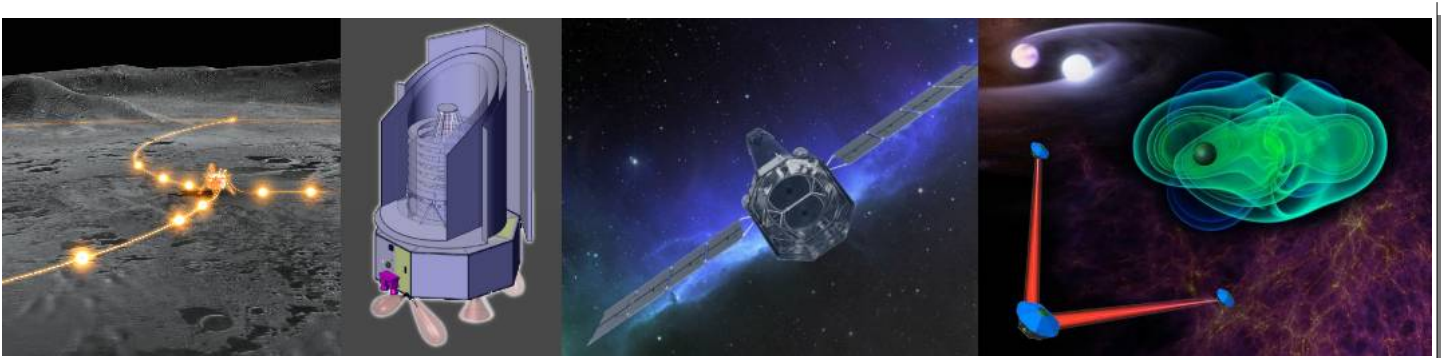

*Figure 24: Possible future missions, from left to right: a low-frequency radio telescope on the Moon, the EChO mission for exoplanet characterisation, the ATHENA X-ray mission and the eLISA/NGO mission for low-frequency gravitational wave sources. Credit: RU Nijmegen/ESA/NGO Project*

In the optical-infrared, developments after the E-ELT strongly point towards arrays of telescopes, linked by fibres, or with individual detectors. The construction of such large telescope arrays will require precursor projects on a smaller scale to build up expertise in this sensor-array approach.

Fundamental research and development may also lead to even longer term goals, such as a low- frequency radio interferometer on the Moon. Radio frequencies below 10 MHz are inaccessible from Earth due to the opaqueness of the ionosphere. A LOFAR-like interferometer on the far side of the Moon is a long term goal to open up this window in the electromagnetic spectrum. Enabling technology includes low-weight antennas, space-based radio interferometry and high-speed, low-power data processing.

## 11 Concluding remarks

Over the last decade the Dutch astronomical community has been able to maintain and even strengthen its leadership position in global astronomy, despite growing competition. It has done so through a nationally-coordinated research and technology development programme, by rejuvenating research staff, by a large-scale investment in modern instrumentation development and establishing better links with industry within the Netherlands and Europe, and by seizing opportunities when they arose. To maintain the leadership position of the Netherlands in global astronomy will require at least as much effort as in the last decade. The recently-completed rejuvenation of the research staff and the investments already made by the Dutch government in, particularly, NOVA, the E-ELT, ALMA, LOFAR and Herschel put Dutch astronomy in an almost-ideal starting position. Excellence does have its price, but through keen choices the investment in astronomy in the Netherlands is, and is foreseen to remain, at a level on par with that of neighbouring European countries. A key condition is the continued investment by the government in fundamental research through viable grants competitions, and the engagement by the astronomical community of high-tech industry in the projects of the future.

Realizing the goals set out in this strategic plan ensures that the Netherlands continues to explore and enjoy the grand view and challenge the Universe offers from front-row seats.

# List of Acronyms

| | |
|---|---|
| ALLEGRO | ALMA Local Expertise Group |
| ALMA | Atacama Large Millimeter/submillimeter Array |
| APERTIF | Focal Plane Array for the WSRT |
| ASPERA | European Roadmap for astroparticle physics |
| ASTRON | Netherlands Institute for Radio Astronomy |
| ASTRONET | A strategic planning mechanism for all of European astronomy |
| AstroWISE | European software environment for astronomical databases |
| ATHENA | Next generation ESA X-ray satellite |
| BSIK | Grant scheme for Large Infrastructure |
| CERN | European Research Centre for High-Energy Physics |
| Chandra | NASA X-ray Observatory |
| CTA | Čerenkov Telescope Array |
| EChO | Exoplanet Characterisation Observatory |
| E-ELT | European Extremely Large Telescope |
| EHT | Event Horizon Telescope |
| ELI | Ministry of Economic Affairs, Agriculture and Innovation |
| eLISA | European Laser Inferometer Space Antenna |
| EMIR | Spanish Infrared MOS instrument for Grantecan |
| ERC | European Research Council |
| ERC-AG | Advanced Grant of the ERC |
| ERC-StG | Starting Grant of the ERC |
| ERC-SyG | Synergy Grant of the ERC |
| ESA | European Space Agency |
| ESFRI | European Strategy for Research Infrastructure |
| ESO | European Southern Observatory |
| EU | European Union |
| Euclid | ESA cosmology space mission |
| Fermi | NASA γ-ray satellite |
| FOM | Physics division of NWO |
| Herschel | ESA's Herschel Space Observatory |
| HIFI | Heterodyne Instrument for Far Infrared |
| JCMT | James Clerk Maxwell Telescope |
| ICT | Information and Communication Technology |
| ILT | International LOFAR Telescope |
| ING | Isaac Newton Group of Telescopes on the island of La Palma |
| INTEGRAL | ESA γ-ray satellite |
| IR | Infrared wavelength regime |
| JAXA | Japanese Space Agency |
| JIVE | Joint Institute for VLBI in Europe |
| JWST | James Webb Space Telescope |
| LHC | Large Hadron Collider at CERN |
| LIGO | Laser Interferometer Gravitational wave Observatory |
| LOFAR | Low Frequency Array |
| LOFT | Large Orbiter for X-ray Timing |
| LRX | Lunar Radio eXplorer |
| LSST | Large Synoptic Survey Telescope |
| LU | Leiden University |
| MIRI | Mid-infrared imaging spectrograph on JWST |
| MKB | Small- and Medium-sized Businesses |
| MOS | Multi-object spectrograph |
| MUSE | Multi-object Unit Spectroscopic Explorer |
| NCA | National Committee for Astronomy |
| NGO | New Gravitational Wave Observatory (formerly eLISA) |
| NIC | NOVA Information Centre |
| NIKHEF | National Institute for Particle and High-Energy Physics |
| NOVA | Netherlands Research School for Astronomy |
| NWO | Netherlands Organisation for Scientific Research |
| NWO-ALW | Division of Earth and Life Sciences of NWO |
| NWO-CW | Division of Chemical Sciences of NWO |
| NWO-EW | Division of Physical Sciences of NWO |
| OCW | Ministry of Education, Culture and Science |
| O/IR | Optical Infra-red wavelength regime |
| O/IR Lab | NOVA instrumentation laboratory |
| OmegaCAM | Wide-field imager for the VST |
| OmegaCEN | Data centre for OmegaCAM |
| PTF2 | Palomar Transient Factory, 2nd generation |
| R&D | Research and development |
| RU | Radboud University Nijmegen |
| RUG | University of Groningen |
| SAFARI | Infrared spectrograph on board SPICA |
| SCUBA-2 | Submillimetre survey camera on JCMT |
| SKA | Square Kilometre Array |
| SPICA | Japanese-European infrared satellite |
| SRON | Netherlands Institute for Space Research |
| Swift | US-UK X-ray satellite for transients |
| TARGET | Large-scale ICT project at RUG |
| VIRGO | European gravitational wave telescope |
| VISTA | Visual Infrared Survey Telescope |
| VLBI | Very Long Baseline Interferometry |
| VLT | Very Large Telescope |
| VLTI | Very Large Telescope Interferometer |
| VST | VLT Survey Telescope |
| WSRT | Westerbork Synthesis Radio Telescope |
| UvA | University of Amsterdam |
| UK | United Kingdom |
| US | United States of America |
| UV | Ultra violet |
| UU | Utrecht University |
| XMM-Newton | ESA's current X-ray satellite |
| X-shooter | VLT optical-infrared spectrograph |

Back cover: a sample of PhD theses in astronomy over the last decade, illustrating the science harvest in astronomy and technology development in the Netherlands

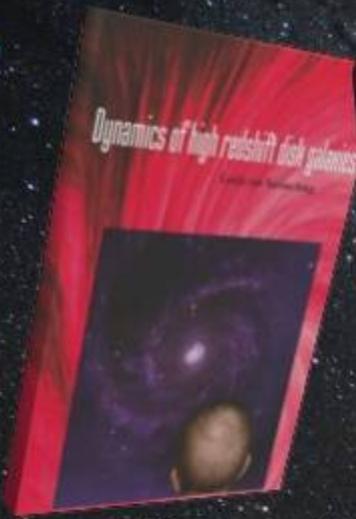
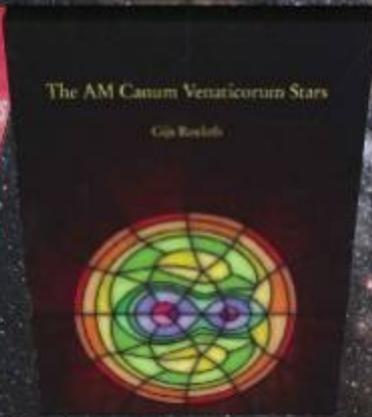
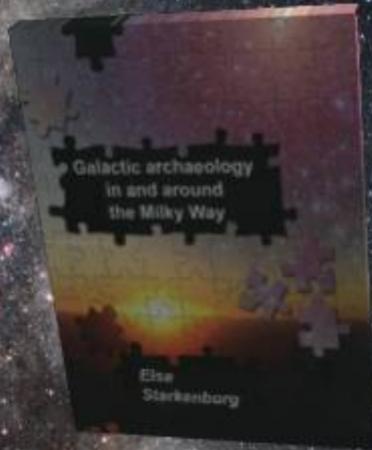
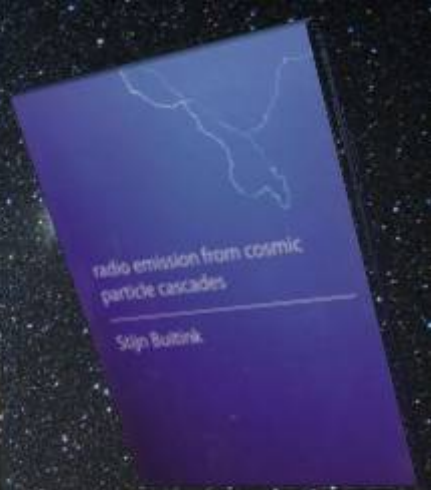
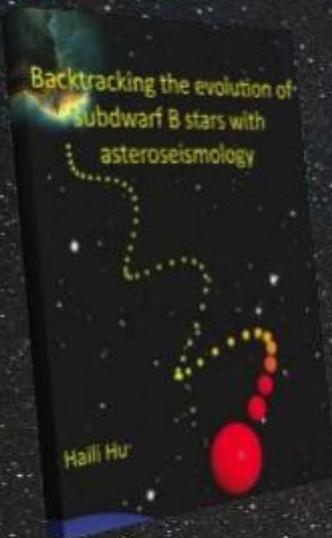
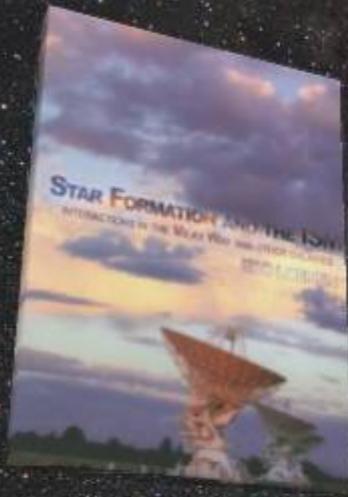
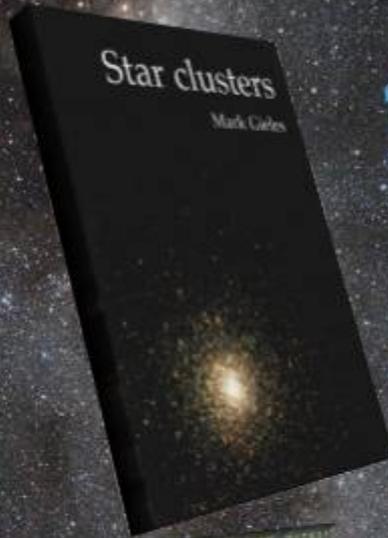
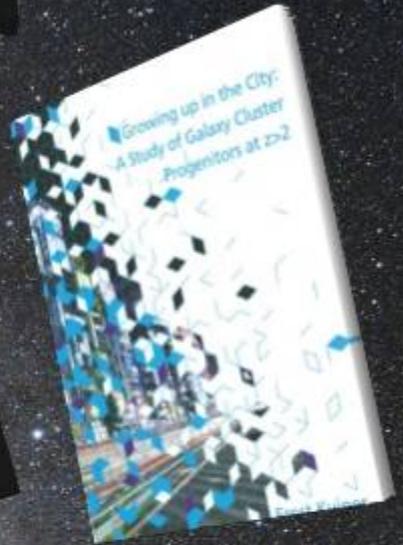
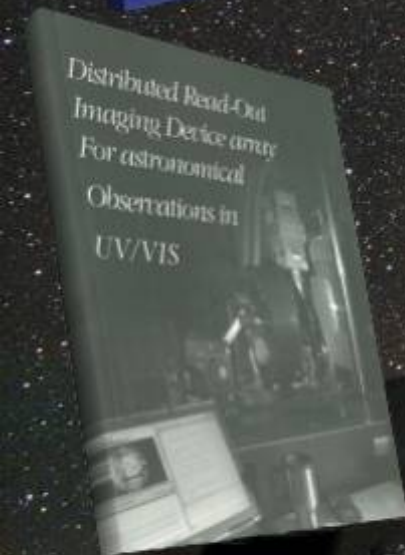
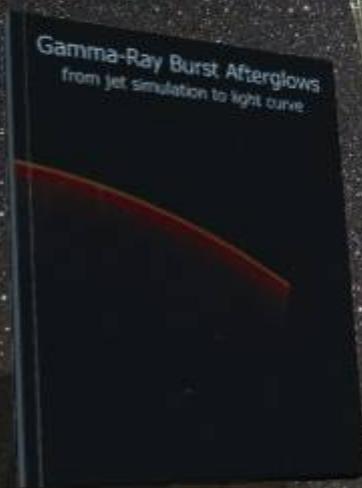
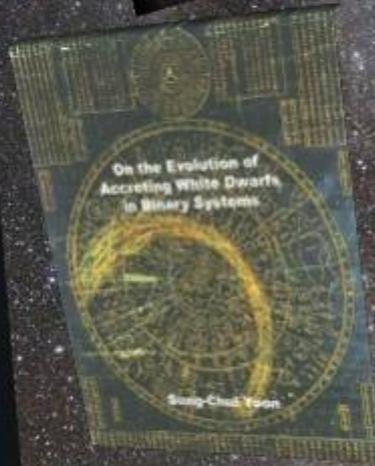
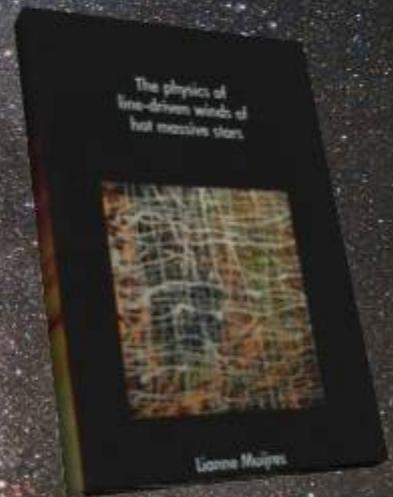
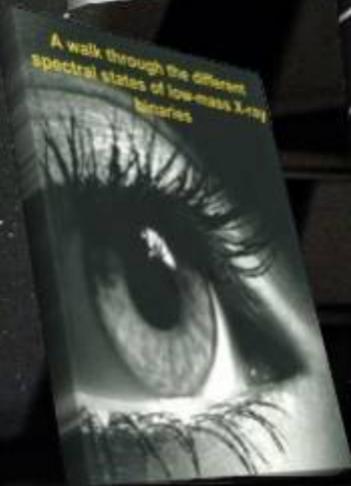
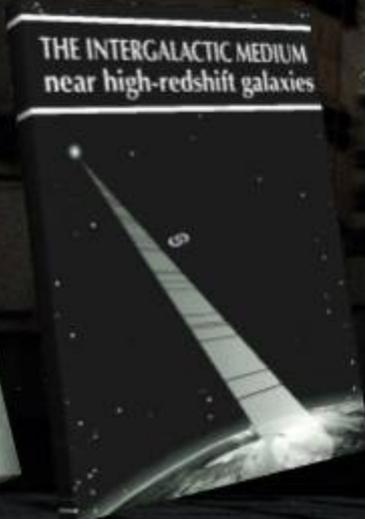
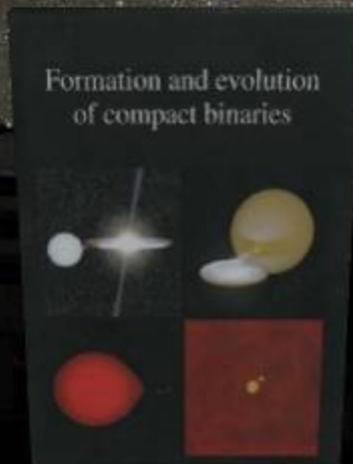
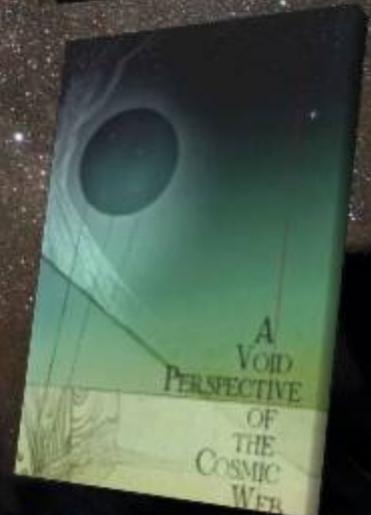